\documentstyle[12pt]{article}
\catcode`\@=11
\def\newsymbol#1#2#3#4#5{\let\next@\relax%
 \ifnum#2=\@ne\else%
 \ifnum#2=\tw@\let\next@\msyfam@\fi\fi%
 \mathchardef#1="#3\next@#4#5}
\def\mathhexbox@#1#2#3{\relax%
 \ifmmode\mathpalette{}{\m@th\mnnathchar"#1#2#3}
 \else\leavevmode\hbox{$\m@th\mathchar"#1#2#3$}\fi}
\font\tenmsy=msbm10
\font\sevenmsy=msbm7
\font\fivemsy=msbm5
\newfam\msyfam
\textfont\msyfam=\tenmsy
\scriptfont\msyfam=\sevenmsy
\scriptscriptfont\msyfam=\fivemsy
\edef\msyfam@{\hexnumber@\msyfam}
\def\Bbb#1{\fam\msyfam\relax#1}
\catcode`\@=\active

\newtheorem{theorem}{Theorem}[section]
\newtheorem{proposition}[theorem]{Proposition}
\newtheorem{lemma}[theorem]{Lemma}
\newtheorem{corollary}[theorem]{Corollary}

\newtheorem{example}[theorem]{Example}
\newtheorem{remark}[theorem]{Remark}

\newcommand{\eq}[1]{\begin{equation}\label{#1}}
\newcommand{\en}{\end{equation}}
\newcommand{\eqn}{\end{eqnarray*}}

\newcommand{\av}{A_\vp}
\newcommand{\bv}{B_\vp}
\newcommand{\ooo}{\sum_{\mu=1,2,3}}
\newcommand{\mmm}{\frac{\vp(k)}{\sqrt{2\opf(k)}} e^{-ikx}\ekj }
\newcommand{\mmmm}{\frac{\vp(k)}{\sqrt{2\opf(k)}} e^{-ikx}(-ik\times \ekj)}

\newcommand{\ekj}{e(k,j)}
\newcommand{\wk}{K^{(2)}}
\newcommand{\kk}{K^{(1)}}
\newcommand{\wwk}{\widetilde{K}}

\newcommand{\proof}{{\noindent \it Proof:\ }}
\newcommand{\qed}{\hfill $\Box$\par\medskip}

\newcommand{\BR}{{{\Bbb R}^3}}
\newcommand{\BD}{{{\Bbb R}^3}}
\newcommand{\BT}{{{\Bbb R}^d}}
\newcommand{\bk}{\BT\setminus K}

\newcommand{\bi}{\begin{description}}
\newcommand{\eni}{\end{description}}

\newcommand{\betai}{\beta_{\rm int}}

\newcommand{\bh}{\ov{H}_0}
\newcommand{\bhb}{\ov{\gsbb}}
\newcommand{\tgsb}{T_{\rm GSB}}

\newcommand{\gsb}{H_{\rm GSB}}
\newcommand{\gsbb}{H_{\rm GSB,0}}
\newcommand{\gsbbb}{H_{\rm GSB, I}}
\newcommand{\tpf}{{T_{\rm PF}}}

\newcommand{\ba}{\ov{A}}

\newcommand{\cD}{{\cal D}}
\newcommand{\DD}{1\otimes N^\han}

\newcommand{\CC}{{{\Bbb C}}}
\newcommand{\tr}{{\rm Tr}}

\newcommand{\RR}{{\Bbb R}}
\newcommand{\WW}{{\cal W}}

\newcommand{\limn}{\lim_{n\rightarrow\infty}}
\newcommand{\limm}{\lim_{m\rightarrow\infty}}
\newcommand{\limt}{\lim_{t\rightarrow\infty}}

\newcommand{\kak}[1]{(\ref{#1})}

\newcommand{\vp}{{\hat\varphi}}

\newcommand{\am}{\ov{A_M}}

\newcommand{\LW}{L^2(\BD\!\!\times\!\!\{1,2\})}
\newcommand{\LR}{{L^2(\BR)}}
\newcommand{\LT}{{L^2(\RR^d)}}

\newcommand{\J}{K_0}
\newcommand{\JJ}{(\hq-z)\f}

\newcommand{\fff}{{{\cal F}_{\rm b}}}
\newcommand{\fffc}{{\oplus^2{\cal F}_{\rm b}}}
\newcommand{\fffa}{{{\cal F}_{\rm f}}}
\newcommand{\hhhd}{\hhh_{\rm D}}
\newcommand{\T}{{T_{\gr}}}
\newcommand{\TT}{{T^\ast_{\gr}}}
\newcommand{\kap}{{\kappa_{\gr}}}

\renewcommand{\j}{{\bf j}}

\newcommand{\hhhm}{\hhh_-}
\newcommand{\hhhp}{\hhh_+}
\newcommand{\hhhh}{\hhh_{\rm d}}
\newcommand{\pgn}{\psi_{\gamma,n}}
\newcommand{\II}{i_1\leq \cdots\leq i_n}
\newcommand{\egn}{E_{\gamma,n}}
\newcommand{\eb}{E_{\rm bin}}
\newcommand{\pgm}{\psi_{\gamma,-}}
\newcommand{\pgp}{\psi_{\gamma,+}}
\newcommand{\bgn}{b_{\gamma,n}}
\newcommand{\bgm}{b_{\gamma,-}}
\newcommand{\bgp}{b_{\gamma,+}}
\newcommand{\bdd}{b^\ast}

\newcommand{\fffo}{{{\cal F}_\omega}}
\newcommand{\hhh}{{\cal H}}
\newcommand{\ffff}{{\cal F}_{\rm fin}}
\newcommand{\FFF}{{\cal F}}
\newcommand{\FFFg}{{\cal F}_{\rm GSB}}
\newcommand{\FFFp}{{\cal F}_{\rm PF}}

\newcommand{\non}{\nonumber}

\renewcommand{\k}{\kap_j}

\newcommand{\qqq}{{\cal P}_\epsilon}
\newcommand{\epp}{\epsilon}
\newcommand{\re}{\rho_\epsilon}

\newcommand{\qq}{{{\cal P}_\Omega}}

\newcommand{\is}{\inf\sigma}

\newcommand{\lll}{\mu=1,2,3}

\newcommand{\f}{^{-1}}

\renewcommand{\P}{P_{\rm field}}

\newcommand{\el}{\end{lemma}}
\newcommand{\hp}{{{H_{\rm p}}}}
\newcommand{\hpp}{{{h_{\rm p}}}}

\newcommand{\lk}{\left(}
\newcommand{\rk}{\right)}
\newcommand{\lkk}{\left\{}
\newcommand{\rkk}{\right\}}

\newcommand{\wick}[1]{\,\ldd{#1}\rdd\,}
\newcommand{\ldd}{:\!\!}
\newcommand{\rdd}{\!\!:}

\newcommand{\bl}[1]{\begin{lemma}\label{#1}}
\newcommand{\bt}[1]{\begin{theorem}\label{#1}}
\newcommand{\et}{\end{theorem}}
\newcommand{\bp}[1]{\begin{proposition}\label{#1}}
\newcommand{\ep}{\end{proposition}}
\newcommand{\bc}[1]{\begin{corollary}\label{#1}}
\newcommand{\ec}{\end{corollary}}

\newcommand{\add}{a^{\dagger}}

\newcommand{\ad}{a^\dagger}

\newcommand{\ass}{a^\sharp}

\newcommand{\ov}[1]{\overline{#1}}

\newcommand{\ob}{{\Omega}}
\newcommand{\obb}{{\Omega}}

\newcommand{\hd}{H_{\rm D}}
\newcommand{\cd}{H_{\rm CD}}

\newcommand{\hf}{H_{\rm f}}

\newcommand{\pf}{{H_{\rm PF}}}
\newcommand{\pfz}{{H_{\rm PF}^0}}
\newcommand{\pff}{{H_{\rm PF,0}}}
\newcommand{\pfs}{{H_{\rm PF}^{\rm sl}}}

\newcommand{\hq}{{H_{\rm q}}}
\newcommand{\ogsb}{{\omega_{\rm GSB}}}
\newcommand{\opf}{{\omega_{\rm PF}}}

\newcommand{\dg}[1]{d\Gamma(#1)}
\newcommand{\dgd}{d\Gamma_{\rm D}}

\newcommand{\gr}{\varphi_{\rm g}}

\newcommand{\han}{{1/2}}

\newcommand{\ci}{c_{\rm int}}
\newcommand{\cex}{c_{\rm exp}}

\newcommand{\hi}{{H_{\rm I}}}
\newcommand{\hicd}{H_{\rm rad}}
\newcommand{\hiicd}{H_{\rm Coulomb}}
\newcommand{\hip}{H_{\rm PF, I}}
\newcommand{\hii}{H_{\rm II}}

\newcommand{\pg}{P_H}
\newcommand{\pgf}{P_\pf}
\newcommand{\pgq}{P_\hq}

\newcommand{\pa}{P_A}
\newcommand{\dr}{{\rm dim}}
\newcommand{\ma}{{\rm m}}

\newcommand{\jj}{\sum_{j=1,2}}

\newcommand{\PP}[1]{e^{-it{#1}}}
\renewcommand{\P}[1]{e^{it{#1}}}
\newcommand{\QQ}[1]{e^{-is{#1}}}
\newcommand{\Q}[1]{e^{is{#1}}}

\newcommand{\bj}{{\bf j}(x)}

\makeatletter
\@addtoreset{equation}{section}
\makeatother

\title
{Multiplicity of ground states  in quantum field models:
applications of asymptotic fields}
\author{
Fumio Hiroshima\thanks{
Department of Mathematics and Physics,
Setsunan University,   572-8508, Osaka, Japan, \ \ \ \ \ \ \ \
\ \ \ \ \ e-mail: hiroshima$@$mpg.setsunan.ac.jp}
\thanks{This work is partialy supported by Grant-in-Aid  for Science
Research (C) 1554019
from MEXT.
}
}
\date{\today}
\begin{document}

\setlength{\baselineskip}{18pt}
\maketitle

\begin{abstract}
The  ground states of an abstract  model in  quantum field
theory are  investigated.
By means of the asymptotic field theory, we give
a necessary and sufficient condition for  that the expectation value of the
number operator of ground states is finite, 
from which we obtain 
a wide-usable  method to estimate
an upper bound of the multiplicity of  ground states.
Ground states of massless GSB models and the Pauli-Fierz model with spin $\han$
are investigated as examples.
\end{abstract}
{\footnotesize
\tableofcontents
}
\newpage

\section{Preliminaries}
\subsection{Boson Fock spaces}
\label{s12}
Let $\WW$ be a Hilbert space over $\CC$ with a conjugation $\bar{ \ }$.
 The boson Fock space $\fff$ over $\WW$ is defined by
\begin{eqnarray*}
\fff&=&\fff(\WW):=\bigoplus_{n=0}^\infty [\otimes_s^n \WW]\\
&=&\{\Psi=\{\Psi^{(n)}\}_{n=0}^\infty| \Psi^{(n)}\in \otimes_s^n \WW,
\|\Psi\|_{\fff}^2:= \sum_{n=0}^\infty \|\Psi^{(n)}\|^2_{\otimes^n
\WW}<\infty\},
\end{eqnarray*}
where $\otimes_s^n \WW$ denotes the $n$-fold symmetric tensor product of
$\WW$
with
$\otimes_s^0 \WW:=\CC$.
In this paper $(f,g)_{\cal K}$ and $\|f\|_{\cal K}$
denote the scalar product and the norm on  Hilbert space ${\cal K}$ over
$\CC$, respectively,
where $(f,g)_{\cal K}$ is linear in $g$ and antilinear in $f$.
Unless  confusions arise
we omit ${\cal K}$ of $(\cdot,\cdot)_{\cal K}$ and $\|\cdot\|_{\cal K}$.
$D(T)$ denotes the domain of operator $T$.
Moreover, for a bounded operator $S$,  we denote its operator norm by
$\|S\|$.

Fock vacuum $\obb\in\fff$ is given by
$$\obb=\{1,0,0,...\}.$$
The finite particle subspace of $\fff$ is defined by
$$\ffff:=
\{\Psi=\{\Psi^{(n)}\}_{n=0}^\infty\in\fff| \Psi^{(m)}=0\mbox{ for all }
m\geq n \mbox{ with some } n\}.$$
It is known that $\ffff$ is dense in $\fff$.
The creation operator $\add(f):\fff\rightarrow \fff$ with test function
$f\in
\WW$ is
the  densely defined  linear operator in $\fff$ defined by
\begin{eqnarray*}
&&(\add(f)\Psi)^{(0)}=0,\\
&&(\add(f)\Psi)^{(n)}=\sqrt n S_n(f\otimes\Psi^{(n-1)}),\ \ \ n\geq 1,
\end{eqnarray*}
where $S_n$ is the symmetrization operator on $\otimes ^n \WW$, i.e.,
$S_n[\otimes^n\WW]=\otimes_s^n\WW$.
The annihilation operator $a(f),f\in \WW$,
is defined by
$$a(f)=(\add(\ov f))^\ast\lceil_{\ffff}.$$
Since it is seen that $a(f)$ and $\add(f)$  are closable operators,
their  closures are denoted by the same symbols, respectively.
Note that $\ass(f)$ ($\ass=a \mbox{ or }  \add$) is linear in $f$.
On $\ffff$ the annihilation operator and the creation operator  obey
canonical commutation relations,
\begin{eqnarray*}
&&[a(f),\add(g)]=(\ov f, g)_\WW,\\
&&[a(f),a(g)]=0,\\
&&[\add(f),\add(g)]=0,
\end{eqnarray*}
where $[A,  B]:=AB-BA$. Define
$$\ffff^D:=\mbox{the linear hull of } \{\add(f_1)\cdots \add(f_n)\ob, \ob|
f_j\in D, j=1,...,,n\geq 1\}. $$
Let $S$ be a self-adjoint operator acting in $\WW$.
The second quantization of $S$,
$$d\Gamma(S):\fff\rightarrow \fff,$$
is defined by
$$
d\Gamma(S):=
\bigoplus_{n=0}^\infty \lk \sum_{j=1}^n \underbrace{1 \otimes\cdots\otimes
\stackrel{j}{S}
\otimes \cdots \otimes 1}_n\rk,$$
with
$$
D(\dg S):=\ffff^{D(S)}.
$$
Here we define
$$(\dg S\Psi)^{(0)}:=0.$$
In particular it follows that
 \eq{omega}
\dg S{\obb}=0.
\en
Note that
\eq{sae}
\dg S\add(f_1)\cdots \add(f_n)\ob=
\sum_{j=1}^n \add(f_1)\cdots \add(S f_j) \cdots \add(f_n)\ob.
\en 
From \kak{sae} it follows that, for $f\in D(S)$,
\begin{eqnarray}
&& \label{dg3}
[\dg S, a(f)]=-a(Sf),\\
&& \label{dg4}
[\dg S, \add (f)]= \add (Sf)
\end{eqnarray}
on $\ffff^{D(S)}$.
It is known that $\dg S$ is essentially self-adjoint.
The self-adjoint extension of $\dg S$ is denoted by the same symbol $\dg S$.
It can be seen that unitary operator $e^{it\dg S}$ acts as 
$$e^{it\dg S} \add(f_1)\cdots \add(f_n)\ob=
\add(e^{itS} f_1)\cdots \add(e^{itS}f_n)\ob.$$
Thus we see that 
that
\begin{eqnarray}
&& \label{dg1}
\P{\dg S} a(f) \PP {\dg S}=a(\PP S f),\\
&&\label{dg2}
 \P{\dg S} \add (f) \PP {\dg S}=\add (\P S f)
\end{eqnarray}
on $\ffff$. 
For a self-adjoint operator $T$, we write its spectrum (resp. essential
spectrum,
point spectrum)
as  $\sigma(T)$ (resp. $\sigma_{\rm ess}(T)$, $\sigma_{\rm p}(T)$).
It is known that
\begin{eqnarray}
\label{11}
&&\sigma(\dg S)=\ov{\{0\}\bigcup\cup_{n=1}^\infty\lkk\left.
\sum_{j=1}^n\lambda_j\right|\lambda_j\in \sigma(S),j=1,...,n\rkk},\\
&&
\label{22}\sigma_{\rm p}(\dg S)=\{0\}\bigcup\cup_{n=1}^\infty\lkk\left.
\sum_{j=1}^n\lambda_j\right|\lambda_j\in \sigma_{\rm p}(S),j=1,...,n\rkk,
\end{eqnarray}
where $\ov{\{\cdots\}}$ denotes the closure of set $\{\cdots\}$.
The second quantization of the identity operator $1$ on ${\cal W}$,
$d\Gamma(1)$,
is referred to as the number operator, which is  written as
$$N:=\dg 1.$$
We note that
$$D(N^k)=\{\Psi=\{\Psi^{(n)}\}_{n=0}^\infty|  \sum_{n=0}^\infty
n^{2k}\|\Psi^{(n)}\|^2<\infty\}.$$
From \kak{11} and \kak{22}
it follows that
$$\sigma(N)=\sigma_{\rm p}(N)={\Bbb N}\cup\{0\}.$$

\subsection{Abstract interaction systems}
Let
$\hhh$ be a Hilbert space.
A Hilbert space for an  abstract coupled system is given by
$$\FFF:=\hhh\otimes\fff,$$
and a decoupled Hamiltonian $H_0$ acting in $\FFF$ is  of the form
$$
H_0=A\otimes 1+1\otimes \dg S.$$
Assumptions {\bf (A1)} and {\bf (A2)} are as follows.
\bi
\item[(A1)]
 Operator $A$ is a self-adjoint operator acting in $\hhh$, and bounded from
below.
\item[(A2)] Operator $S$ is a nonnegative self-adjoint operator acting in
${\cal W}$.
\eni
Total Hamiltonians under consideration are of the form
\eq{R}
H=H_0+g\hi,
\en
where
$g\in\RR$ denotes a coupling constant and $\hi$ a symmetric operator.
Assumption {\bf (A3)} is as follows.
\bi
\item[(A3)]
$\hi$ is $H_0$-bounded  with
$$\|\hi\Psi\|\leq a\| H_0\Psi\|+b\|\Psi\|,\ \ \ \Psi\in D(H_0),$$
where $a$ and $b$ are nonnegative constants.
\eni
Under {\bf (A3)}, by the Kato-Rellich theorem,
$H$  is self-adjoint on $D(H_0)$ and
bounded from below for $g$ with $|g|<1/a$.
Moreover $H$ is essentially self-adjoint on any core of $H_0$.
The bottom of $\sigma (H)$ is denoted by
$$E(H):=\is (H),$$
which is referred to as the ground state energy of $H$.
If an eigenvector  $\Psi$ associated with $E(H)$  exists,
i.e.,
$$H\Psi=E(H)\Psi, $$
then $\Psi$ is called a ground state of  $H$.
Let
$E_T(B)$ be the spectral projection of self-adjoint operator $T$ onto a
Borel set $B\subset \RR$. We set 
$$P_T:=E_T(\{E(T)\}).$$
Then 
$P_H$ 
denotes the projection onto the subspace  spanned by ground states of $H$.
The dimension of $P_H \FFF$
is called the multiplicity of ground states of $H$, and it is denoted by
$$\ma(H):={\rm dim}\ P_H \FFF.$$
If  $\ma(H)=1$,
then we call that the ground state of $H$ is unique.

\subsection{Expectation values of the number operator}
\label{plo}
For Hamiltonians like as \kak{R},
the existence  of a ground state $\gr$
such that
\eq{mass}
\gr\in D(1\otimes N^\han)
\en
has been shown by many authors, e.g., \cite{ah1,  bfs1, bfs3, ge, gll, h8,
sp}.
Conversely,  if $\gr$ exists, little attention, however,  has been given to
investigate whether
\kak{mass} holds or not.
Then the first task in this paper is to 
give  a necessary and sufficient condition for 
\eq{ko}
P_H\FFF\subset D(1\otimes N^\han).
\en
As we will see later, to show \kak{ko} is also
the primary problem in estimating an upper bound of
$\ma(H)$.

\subsection{Massive and massless cases}
Typical examples of Hilbert space ${\cal W}$ and 
nonnegative self-adjoint operator $S$ are 
\begin{eqnarray}
&&\label{ko1}
{\cal W}=\LT,\\
&&\label{ko2}
S= \mbox{ the multiplication operator by }
\omega_\nu(k):=\sqrt{|k|^2+\nu^2}.
\end{eqnarray}
In the case of $\nu>0$ (resp. $\nu=0$), a model is referred to as
a {\it massive} (resp. {\it massless}) model.
Note that under {\bf (A1)} and {\bf (A3)},
\eq{ty}
D(H)=D(H_0)=D(A\otimes 1)\cap D(1\otimes \dg{\omega_\nu}).
\en
In a  massive  case,  one can  see that \kak{ko} is always satisfied.
Actually in a  massive case, we have
$D(\dg{\omega_\nu})\subset D( N)$ and
$$\frac{1}{\nu} \| \dg{\omega_\nu}\Psi\| \geq \|N\Psi\|,\ \ \ \Psi\in
D(\dg{\omega_\nu}).$$
Together with \kak{ty} we obtain that
$$P_H\FFF\subset D(H)\subset D(1\otimes \dg{\omega_\nu})\subset D(1\otimes
N)\subset
D(1\otimes N^\han).$$
Hence \kak{ko} follows.
Kernel $a(k)$ of $a(f)$, $f\in\LT$, is defined for each $k\in\BT$ as
$$\lk a(k)\Psi\rk^{(n)}(k_1,...,k_n)=\sqrt{n+1}\Psi^{(n+1)}(k,k_1,...,k_n)$$
and
$$\lk a(f)\Psi\rk^{(n)}=\int f(k) (a(k)\Psi)^{(n)} dk$$
for  $\Psi\in \ffff^{C_0^\infty(\BT)}$, 
and it is directly seen that 
\eq{masss}
\int_\BT \|a(k) \Psi\|^2 dk =  \|N^\han \Psi\|^2,\ \ \ \Psi\in \ffff^{C_0^\infty(\BT)}.
\en
From \kak {masss}, 
$a(\cdot)\Psi$ for $\Psi\in D(N^\han)$ can be  defined
as an $\fff$-valued $L^2$ function on $\BT$  by
$$a(\cdot)\Psi:=s-\lim_{m\rightarrow \infty}a(\cdot)\Psi_m\ \ \ in\  \
L^2(\BT;\fff),$$
where $s-\lim_{m\rightarrow \infty}$ denotes the strong limit in  
$L^2(\BT;\fff)$ and 
 sequence $\Psi_m\in
\ffff^{C_0^\infty(\BT)}$ is 
such that
$\Psi_m\rightarrow \Psi$ and $N^\han\Psi_m\rightarrow N^\han\Psi$ strongly as 
$m\rightarrow \infty$.
By  an {\it informal} calculation, it can be derived {\it pointwise}  that
\eq{pl2}
(1\otimes a(k))\gr= g(H-E(H)+\omega(k))^{-1}[\hi, 1\otimes a(k)]\gr. 
\en
Note that at least we have to assume $\gr\in D(\DD)$ for  \kak{pl2} to make a sense, 
and the right-hand side of \kak{pl2} is also delicate.  
See e.g., \cite[Lemma 2.6]{sl} 
and \cite[p.170, {\it Conclusion }]{fr2} for this point.  
For massive cases, 
$(1\otimes a(\cdot))\gr$   is  well defined as
an $\FFF$-valued $L^2$ function on $\BT$, since  
$\gr\in D(\DD)$,
but of course it does not make  sense  pointwise.
From \kak{pl2}  and \kak{masss} it follows that
\eq{pl}
\|(1\otimes N^\han)\gr\|^2
=
g^2 \int_\BT  \| (H-E(H)+\omega(k))^{-1}[\hi, 1\otimes a(k)]\gr\|^2 dk.
\en
We may say  under some conditions that 
\begin{eqnarray*}
&&\gr\in D(\DD) \mbox{ and }
\int_\BT  \| (H-E(H)+\omega(k))^{-1}[\hi, 1\otimes a(k)]\gr\|^2 dk<\infty
\non \\
&&\label{pk}
\Longrightarrow \|(\DD)\gr\|^2=
g^2\int_\BT  \| (H-E(H)+\omega(k))^{-1}[\hi, 1\otimes a(k)]\gr\|^2 dk.
\end{eqnarray*}
Although \kak{pl} has been  applied 
to study $\|(1\otimes N^\han)\gr\|$ by many authors,
it must be noted again that \kak{pl} is derived from 
{\it informal} formula \kak{pl2}. 

We are most interested in analysis of ground states for massless cases. 
In this case $\gr\in D(1\otimes N^\han)$ is not clear, 
and it is also not clear a priori that  $(1\otimes a(k))\gr$ makes a sense. 
Then it is uncertain that  identity \kak{pl2} holds true 
for massless cases in some sense. 
Furthermore the fact that the right-hand side of \kak{pl} is finite 
does not play a role in a criterion for whether $\gr\in D(\DD)$ or not, 
since,  at least,  we have to assume $\gr\in D(\DD)$ in \kak{pl2}.

Because of the tedious argument involved in establishing \kak{pl2}
pointwise,
a quite different method  is taken to show  \kak{pl} in this paper.
We will show under some conditions that 
\eq{km1}
\gr\in D(\DD)\Longleftrightarrow 
\int_\BT  \| (H-E(H)+\omega(k))^{-1}[\hi, 1\otimes a(k)]\gr\|^2 dk<\infty,  
\en 
and \kak{pl} follows  when the right or left-hand side of \kak{km1} holds. 
The method is an application of the fact
that asymptotic annihilation operators vanish arbitrary ground states.
See \kak{ko5}.
As a result,  \kak{pl} and \kak{km1} can be  valid rigorously for both massive and massless
cases
without using \kak{pl2}. As far as we know, this method is new, cf.,
see \cite{ahh, ahh2, hiro0, hiro}.
By means of  \kak{km1} we can find  a condition for 
$\pg\FFF\subset D(1\otimes N^\han)$.

\subsection{Multiplicity}
Generally,  in the case where $E(H)$ is discrete,
the min-max principle \cite{rs4} is available to estimate the  multiplicity
of
ground states.
Actually the ground state energy of
 a  {\it massive} generalized-spin-boson (GSB) model
with a sufficiently weak coupling is discrete.
Hence the min-max principle
can be applied  for this model
\cite{ah1}.
However, for some  typical models,
e.g., massless GSB models, the  Pauli-Fierz model, and the  Nelson model
\cite{ne3}, etc.,
their  ground state energy is the edge of the essential spectrum, namely it
is not discrete. See also \cite{ar14, h12}.
Then the min-max principle does not work at all.

Instead of the min-max principle,
we can apply an infinite dimensional version of the
Perron-Frobenius theorem \cite{gr0, gr1} to show the uniqueness of its
ground
state.
I.e.,
in a Schr\"odinger representation,
\eq{pi}
(\Psi, e^{-tH}\Phi)>0,\ \ \
\Psi\geq 0\  (\not\equiv 0), \ \ \Phi\geq 0\  (\not\equiv 0),
\en
implies $\ma(H)=1$.
Property \kak{pi} is called that $e^{-tH}$ is positivity-improving.
The Perron-Frobenius theorem has been applied for some models,
e.g.,
the Nelson model in \cite{bfs1},
and the spinless Pauli-Fierz model in \cite{h9}.
It is,  however,   for,  e.g., the Pauli-Fierz model with spin $1/2$, $\pf$,
we can not apply the Perron-Frobenius theorem, since, as far as we know,
a suitable representation for $e^{-t\pf }$ to be positivity-improving can
not be constructed.

In this paper,  applying the fact $\pg\FFF\subset D(1\otimes N^\han)$,
we establish a wide-usable  method to estimate
an upper bound of the multiplicity  of  ground states under some conditions.

\subsection{Main results and strategies}
\label{kou}
The main results are  (m1) and (m2).
\bi
\item [(m1)]
We give a necessary and sufficient condition for $P_H \FFF\subset
D(1\otimes N^\han)$.
\item [(m2)]
We prove $\ma(H)\leq \ma(A)$ under some conditions.
\eni
Strategies are as follows.
It is proven that
\begin{eqnarray}
\label{ko3}
\gr\in D(1\otimes N^\han) \Longleftrightarrow
\sum_{m=1}^\infty\|(1\otimes a(e_m))\gr\|^2<\infty,
\end{eqnarray}
where $\{e_m\}_{m=1}^\infty$ is an arbitrary
complete orthonormal system of ${\cal W}$.
When the left or right-hand side of \kak{ko3} holds,
it follows that 
\eq{r}
\sum_{m=1}^\infty\|(1\otimes a(e_m))\gr\|^2=\|(1\otimes N^\han)\gr\|^2.
\en 
Let us define an asymptotic annihilation operator by
\eq{m01}
a_+ (f)\Psi:=s-\limt \PP{H}\P{H_0}(1\otimes a(f)) \PP{H_0}\P H\Psi.
\en
Of course some conditions on $\Psi$ and $f$ are required to show
the existence of $a_+(f)\Psi$.
It is well known \cite{al3, ho1}, however,  that
\kak{m01} exists for an arbitrary ground state of $H$, $\Psi=\gr$,
 and $a_+(f)$ vanishes $\gr$, i.e.,
\eq{ko5}
a_+(f)\gr=0
\en
for some $f$.
 \kak{ko5} is applied for  (m1).
We decompose $a_+(f)\Psi$ as
$$a_+(f)\Psi =(1\otimes a(f))\Psi -g G(f)\Psi $$
with some operator $G(f):\FFF\rightarrow \FFF$.
From \kak{ko5} it follows that
\eq{n2}
(1\otimes a(f) )\gr=gG(f)\gr.
\en
We define the  bounded operator
$\T:{\cal W} \rightarrow \FFF$ by
\eq{n3}
 \T f:=G(f)\gr,\ \ \ f\in {\cal W}.
\en
I.e.,
\eq{ror}
(1\otimes a(f) )\gr=g\T f.
\en
It is seen that $\T$ is an $\FFF$-valued integral operator such that
$$\T f=\int_\BT f(k) \kap (k) dk $$
with some kernel $\kap(k)\in\FFF$.
See \kak{ttt} for details.
Note that
\eq{ro}
\sum_{m=1}^\infty\| \T e_m \|^2=  \tr (\TT \T)=\int_\BT\|\kap(k)\|^2 dk.
\en
Using \kak{ko3}, \kak{ror} and \kak{ro},
we see that
$$
\gr\in D(1\otimes N^\han) \Longleftrightarrow
g^2 \int_\BT\|\kap(k)\|^2 dk<\infty,$$
and by \kak{r}, 
\eq{po2}
\|(1\otimes N^\han)\gr\|^2 = g^2 \int_\BT \|\kap(k)\|^2 dk.
\en
Thus we can obtain that
$$\pg\FFF\subset D(1\otimes N^\han)\Longleftrightarrow
\int\|\kap(k)\|^2dk<\infty
\mbox{ for all } \gr\in \pg\FFF.$$
To show (m2) we apply the method in \cite{hisp2}, by which
we can prove that
$${\rm dim} (\pg\FFF\cap D(1\otimes N^\han) ) \leq \frac{1}{1-\delta(g)}
\ma(A),$$
where
$$\delta(g)=o(g)+\sup_{\gr\in \pg\FFF\cap D(1\otimes N^\han)}
\frac{\|(1\otimes N^\han)\gr\|^2}{\|\gr\|^2}.$$
By \kak{po2} and the fact
$$\lim_{g\rightarrow 0}\sup_{\gr\in \pg\FFF}\frac{\int_\BT\|\kap(k)\|^2
dk}{\|\gr\|^2}<\infty,$$
we see that
$\lim_{g\rightarrow 0}\delta(g)=0$.
Hence for sufficiently small $g$,
$${\rm dim} (\pg\FFF\cap D(1\otimes N^\han) ) \leq  \ma(A)$$
is proven. Together with the fact $\pg\FFF\subset D(1\otimes N^\han)$ under
some conditions,
we get $$\ma(H)={\rm dim} \pg\FFF\leq \ma(A).$$

We organize this paper as follows.

Section 2 is devoted to show $P_H\FFF\subset D(1\otimes N^\han)$.
In Section 3, we estimate the multiplicity of ground states.
In Sections  4, we give examples including
massless GSB models and the Pauli-Fierz model.
Finally in Section 5 we give appendixes.

\section{Equivalent conditions to $P_H\FFF \subset D(1\otimes
N^\han)$}
\subsection{The number operator}
Let $\{e_m\}_{m=1}^\infty$ be a complete orthonormal system of ${\cal W}$.
We define
$$A_M,\ \ \ M=1,2,...,$$
by
$$A_M:=(N+1)^{-\han}\lk\sum_{m=1}^M\add(e_m)a(\ov{e_m})\rk(N+1)^{-\han}.$$
\bl{g1}
We have
\bi
\item[(1)] $A_M$ has a  unique  bounded operator  extension $\ov{A_M}$,
\item[(2)] $\ov{A_M}$ is uniformly bounded in $M$ as $\|\ov{A_M}\|\leq 1$,
\item[(3)] $s-\lim_{M\rightarrow \infty}A_M=N(N+1)^{-1}$.
\eni
\el
\proof
Let us define
$$\fffo:=
\left[
\bigoplus_{n=0}^\infty \lkk\left.
\sum_{\II}^{\rm finite}\alpha_{i_1,...,i_n}
\add(e_{i_1})\cdots \add(e_{i_n}){\obb}\right|
\alpha_{i_1,...,i_n}\in\CC
\rkk\right] \bigcap \ffff.$$
Note that $\fffo$ is dense in $\fff$.
Let
$$
\phi= \add(e_{i_1})\cdots \add(e_{i_n}){\obb},\ \ \
 i_1\leq\cdots\leq {i_n}.
$$
Then
\begin{eqnarray}
A_M\phi&=& \frac{1}{n+1}\sum_{j=1}^n
\add(e_{i_1})\cdots \add(\sum_{m=1}^M(e_m,e_{i_j})e_m)
\cdots \add(e_{i_n}){\obb}\nonumber \\
\label{k1}
&=&\beta_{i_1,...,i_n}(M)\phi,
\end{eqnarray}
where
$$
\beta_{i_1,...,i_n}(M):=
\lkk\begin{array}{ll}\frac{n}{n+1},&i_n\leq M,\\
\frac{n-1}{n+1},& i_{n-1}\leq M<i_n,\\
\ \  \vdots&\ \ \ \ \ \ \ \vdots\\
\frac{1}{n+1},& i_1\leq M<i_2,\\
0,& M<i_1.
\end{array}\right.
$$
Let $\Psi\in\fffo$ be such that
\eq{g2}
\Psi=\sum_{\II}^{\rm finite}\alpha_{i_1,...,i_n}\add(e_{i_1})\cdots
\add(e_{i_n}){\obb}.
\en
We see that
$$\|\Psi\|^2=\sum_{\II}^{\rm finite}|\alpha_{i_1,...,i_n}|^2.$$
From \kak{k1} it follows that
$$A_M\Psi=\sum_{\II}^{\rm finite}\alpha_{i_1,...,i_n} \beta_{i_1,...,i_n}(M)
\add(e_{i_1})\cdots \add(e_{i_n}).$$
Then
\begin{eqnarray*}
0&\leq &  \|A_M\Psi\|^2=\sum_{\II}^{\rm finite}|\alpha_{i_1,...,i_n}|^2
|\beta_{i_1,...,i_n}(M)|^2\\
& \leq &
\sum_{\II}^{\rm finite}|\alpha_{i_1,...,i_n}|^2
\lk \frac{n}{n+1}\rk ^2=\lk\frac{n}{n+1}\rk^2 \|\Psi\|^2.
\end{eqnarray*}
Note that $A_M$ leaves $\otimes_s^n{\cal W}$ invariant.
Hence for an arbitrary $\Psi=\{\Psi^{(n)}\}_{n=0}^\infty \in\fffo$, we have
$$\|A_M\Psi\|^2=
\sum_{n=0}^\infty \|(A_M\Psi)^{(n)}\|^2
=
\sum_{n=0}^\infty \|A_M\Psi^{(n)}\|^2\leq
\sum_{n=0}^\infty \lk \frac{n}{n+1}\rk^2 \|\Psi^{(n)}\|^2\leq \|\Psi\|^2.$$
Since $\fffo$ is dense in $\fff$, (1) and (2) follow.
Let $\Psi$ be as \kak{g2}.
We see that
$$s-\lim_{M\rightarrow \infty} A_M\Psi=\frac{n}{n+1}\Psi.$$
Hence for an arbitrary $\Phi\in\fffo$,
$$s-\lim_{M\rightarrow \infty} A_M\Phi=N(N+1)^{-1}\Phi.$$
For an arbitrary $\Phi\in\fff$ and an arbitrary $\epsilon>0$,
we can choose $\Phi_\epsilon\in\fffo$ such that
$$\|\Phi-\Phi_\epsilon\|<\epsilon.$$
Since $\|\am\|\leq 1$, we obtain that
\begin{eqnarray*}
&& \|\am \Phi-N(N+1)^{-1}\Phi\|\\
&&\leq \|\am \Phi-\am\Phi_\epsilon\|+
\|\am\Phi_\epsilon-N(N+1)^{-1}\Phi_\epsilon\|+
\|N(N+1)^{-1}(\Phi_\epsilon-\Phi)\|\\
&&\leq 2\epsilon+\|\am\Phi_\epsilon-N(N+1)^{-1}\Phi_\epsilon\|.
\end{eqnarray*}
Then
$$\lim_{M\rightarrow\infty} \|\am \Phi-N(N+1)^{-1}\Phi\|<2\epsilon$$
for an arbitrary $\epsilon$.
Thus (3) follows.
\qed
\bl{g4}
Let $\{e_m\}_{m=1}^\infty$ be an arbitrary  complete orthonormal system in
${\cal W}$.
Then
(1) and (2) are  equivalent.
\bi
\item[(1)] $\Psi\in D(N^\han)$.
\item[(2)]
$\Psi\in \cap_{m=1}^\infty D(a(\ov{e_m}))$ and
\eq{as}
\sum_{m=1}^\infty \|a(\ov{e_m})\Psi\|^2<\infty.
\en
\eni
Moreover  when (1) or (2) holds, it follows that
$$\|N^\han\Psi\|^2=\sum_{m=1}^\infty \|a(\ov{e_m})\Psi\|^2.$$
\el
\proof
$(1)\Rightarrow (2)$

We see that $\fffo\subset D(N^\han)$ and
$$e^{-t N^\han }\fffo\subset \fffo,$$
which implies that $\fffo$ is a core of $N^\han$ by \cite[X.49]{rs2}.
Then,  for $\Psi\in D(N^\han)$,
there exists a sequence $\Psi_\epsilon\in\fffo$ such that
$s-\lim_{\epsilon\rightarrow 0}\Psi_\epsilon=\Psi$ and
$s-\lim_{\epsilon\rightarrow 0}N^\han \Psi_\epsilon=N^\han\Psi$.
It is well known that
$$\|a(f)\Phi\|\leq \|f\|\|N^\han \Phi\|,\ \ \ \Phi\in D(N^\han).$$
Hence from the fact $\Psi\in D(N^\han)$, it follows that $\Psi\in
D(a(\ov{e_m}))$.
We have
\begin{eqnarray}
 \sum_{m=1}^M  \|a(\ov{e_m})\Psi_\epsilon\|^2&=&((N+1)^\han\Psi_\epsilon,
A_M (N+1)^\han \Psi_\epsilon)\nonumber \\
&\leq& \|(N+1)^\han\Psi_\epsilon\|^2\nonumber \\
\label{star}
&\leq& \|N^\han\Psi_\epsilon \|^2+\|\Psi_\epsilon \|^2.
\end{eqnarray}
From this it follows that
$a(\ov{e_m})\Psi_\epsilon$ is a Cauchy sequence in $\epsilon$.
Since  $a(\ov{e_m})$ is a closed operator,
$\lim_{\epsilon\rightarrow 0}a(\ov{e_m})\Psi_\epsilon=a(\ov{e_m})\Psi$
follows.
Hence we obtain that,  as $\epsilon\rightarrow 0$ on the both sides of
\kak{star},
$$\sum_{m=1}^M\| a(\ov{e_m})\Psi\|^2 \leq \|N^\han\Psi\|^2+\|\Psi\|^2.$$
Taking  $M\rightarrow \infty$ on the both sides above, we have
$$\sum_{m=1}^\infty \| a(\ov{e_m})\Psi\|^2 \leq
\|N^\han\Psi\|^2+\|\Psi\|^2.$$
Thus the desired results follow.

$(2)\Rightarrow (1)$
We see that
\begin{eqnarray*}
 \sum_{m=1}^\infty \|a(\ov{e_m})\Psi\|^2
&=&
\lim_{M\rightarrow \infty} \sum_{m=1}^M
\sum_{n=0}^\infty (a(\ov{e_m})\Psi^{(n)}, a(\ov{e_m})\Psi^{(n)})\\
&=&
\lim_{M\rightarrow \infty}
\sum_{n=0}^\infty \sum_{m=1}^M(a(\ov{e_m})\Psi^{(n)},
a(\ov{e_m})\Psi^{(n)}).
\end{eqnarray*}
Since
$\sum_{m=1}^M(a(\ov{e_m})\Psi^{(n)}, a(\ov{e_m})\Psi^{(n)})$ is monotonously
increasing as $M\uparrow\infty$ and by \kak{as},
$$\lim_{M\rightarrow \infty}
\sum_{n=0}^\infty \sum_{m=1}^M(a(\ov{e_m})\Psi^{(n)},
a(\ov{e_m})\Psi^{(n)})<\infty,$$
we have by
the Lebesgue monotone convergence theorem and
(3) of Lemma \ref{g1},
\begin{eqnarray*}
&&
\lim_{M\rightarrow \infty}
\sum_{n=0}^\infty \sum_{m=1}^M(a(\ov{e_m})\Psi^{(n)},
a(\ov{e_m})\Psi^{(n)})\\
&&=
\sum_{n=0}^\infty \lim_{M\rightarrow \infty}
\sum_{m=1}^M(a(\ov{e_m})\Psi^{(n)}, a(\ov{e_m})\Psi^{(n)})\\
&& =
\sum_{n=0}^\infty \lim_{M\rightarrow \infty}
((N+1)^\han \Psi^{(n)}, \am (N+1)^\han \Psi^{(n)})\\
&&=
\sum_{n=0}^\infty n
( \Psi^{(n)},  \Psi^{(n)})\\
&&=
\sum_{n=0}^\infty n
\| \Psi^{(n)}\|^2<\infty.
\end{eqnarray*}
This yields that $\Psi\in D(N^\han)$.
\qed

\subsection{Weak commutators}
\label{w1}
In subsections \ref{w1}-\ref{w3},  we consider the case where 
\eq{w5}
{\cal W}=\oplus^D\LT\cong L^2(\BT\times\{1,...,D\})
\en 
and 
\eq{w6}
S=[\omega]
\en 
such that 
$[\omega]:\oplus^D\LT\rightarrow \oplus^D\LT$
is the  multiplication operator defined by 
\eq{ro1}
[\omega] (\oplus^D_{j=1} f_j)=\oplus _{j=1}^D \omega f_j,
\en
with
$$\omega(\cdot):\BT\rightarrow [0,\infty),\ \ \ \
(\omega f)(k)=\omega(k) f(k).
$$
The creation operator and the annihilation operator of $\fff({\cal W})$
are  denoted by
\begin{eqnarray*}
\ass (f,j):=a(0\oplus\cdots \oplus \stackrel{j}{f}
\oplus\cdots\oplus 0), \ \ \ f\in\LT,\ \ \ j=1,...,D,
\end{eqnarray*}
which satisfy on $\ffff$, 
\begin{eqnarray*}
&& [a(f,j),\add(g,j')]=(\bar f, g)\delta_{jj'}, \\
&& [\add(f,j),\add(g,j')]=0,\\
&& [a(f,j),a(g,j')]=0.
\end{eqnarray*}
By \kak{dg1}-\kak{dg4},
\begin{eqnarray}
&& \label{dgg1}
\P{(1\otimes \dg {[\omega]})}
(1\otimes a(f,j) )
\PP {(1\otimes \dg {[\omega]})}=1\otimes a(\PP \omega f,j),\\
&&\label{dgg2}
 \P{(1\otimes \dg {[\omega]})}
(1\otimes \add (f,j) )
\PP {(1\otimes \dg {[\omega]})}=1\otimes \add (\P \omega f,j)
\end{eqnarray}
follow on $\hhh\otimes \ffff$, and
\begin{eqnarray}
&& \label{dgg3}
[1\otimes \dg {[\omega]}, 1\otimes a(f,j)]=-1\otimes a(\omega f,j),\\
&& \label{dgg4}
[1\otimes \dg {[\omega]}, 1\otimes \add (f,j)]= 1\otimes \add (\omega f,j)
\end{eqnarray}
follow 
for $f\in D(\omega)$
on $\hhh\otimes \ffff^{D([\omega])}$.
Let $S$ and $T$ be operators acting in a Hilbert space ${\cal K}$.
We define a quadratic form
$[S, T]_W^{\cal D}$ with a form domain 
$$\cD\subset D(S^\ast)\cap D(S)\cap
D(T^\ast)\cap
D(T)$$ 
by
$$
[S, T]_W^D (\Psi, \Phi):=(S^\ast\Psi, T\Phi)-(T^\ast\Psi, S\Phi),\ \ \
\Psi,\Phi\in \cD. $$
\bp
{well}
(1)
Let $f,f/\sqrt\omega\in \LR$.
Then
\kak{dgg1} and \kak{dgg2} can be extended on
$D(1\otimes \dg{[\omega]})$.
(2)
Let
$\omega f, f/\sqrt\omega\in\LR$.
Then
\begin{eqnarray}
&& \label{to1}
[1\otimes \dg {[\omega]}, 1\otimes a(f,j)]_W^{D(1\otimes
\dg{[\omega]})}(\Psi,\Phi)
=(\Psi, -(1\otimes a(\omega f,j))\Phi),\\
&&
\label{to2}
[1\otimes \dg {[\omega]}, 1\otimes \add(f,j)]_W^{D(1\otimes
\dg{[\omega]})}(\Psi,\Phi)
=(\Psi,  (1\otimes \add (\omega f,j))\Phi).
\end{eqnarray}
\ep
\proof
Let $\Psi\in D(1\otimes \dg{[\omega]})$.
Since $\hhh\otimes \ffff$ is a core of $1\otimes \dg{[\omega]}$,
there exists a sequence $\Psi_\epp$ such that
$\Psi_\epp\rightarrow \Psi$ and $(1\otimes
\dg{[\omega]})\Psi_\epp\rightarrow
(1\otimes \dg{[\omega]})\Psi$ strongly as $\epp\rightarrow 0$.
It follows that 
\begin{eqnarray}
&&
\label{po1}
\P{(1\otimes \dg {[\omega]})}
(1\otimes a(f,j) )
\PP {(1\otimes \dg {[\omega]})}\Psi_\epp
=1\otimes a(\PP \omega f,j)\Psi_\epp,\\
&&\label{pop2}
\P{(1\otimes \dg {[\omega]})}
(1\otimes \add (f,j) )
\PP {(1\otimes \dg {[\omega]})}\Psi_\epp
=1\otimes \add (\P \omega f,j)\Psi_\epp.
\end{eqnarray}
Using well known inequalities
\begin{eqnarray*}
&& \|(1\otimes a(f,j))\Psi\|\leq \|f/\sqrt\omega\|\|(1\otimes
\dg{[\omega]}^\han)\Psi\|,\\
&& \|(1\otimes \add (f,j))\Psi\|\leq
\|f/\sqrt\omega\|\|(1\otimes \dg{[\omega]}^\han)\Psi\|+
\|f\| \|\Psi\|,
\end{eqnarray*}
we see that the both hand sides of \kak{po1} and \kak{pop2} converge
strongly
as $\epsilon\rightarrow 0$. Since $1\otimes \ass(f,j)$ is closed, (1)
follows.

We shall prove (2).
Let $\Psi,\Phi\in \hhh\otimes \ffff^{D([\omega])}$. Then
\begin{eqnarray*}
&& ((1\otimes \dg{[\omega]})\Psi, (1\otimes a(f,j))\Phi)
-((1\otimes \ad(\bar f,j))\Psi, (1\otimes \dg{[\omega]})\Phi)\\
&&\hspace{5cm}
=
(\Psi, -(1\otimes a(f,j))\Phi).
\end{eqnarray*}
Since $\hhh\otimes \ffff^{D([\omega])}$ is a core of $1\otimes
\dg{[\omega]}$,
there exists a sequence $\Psi_\epp$ such that
$ \Psi_\epp\rightarrow \Psi$ and $(1\otimes
\dg{[\omega]})\Psi_\epp\rightarrow
(1\otimes \dg{[\omega]})\Psi$ strongly as $\epp\rightarrow 0$.
By using the closedness of $1\otimes a(f,j)$ and a similar limiting argument
as that of (1),
we obtain \kak{to1}. \kak{to2} can be similarly proven.
\qed

\subsection{Asymptotic fields}
\label{w2}
Define on $D(H)$,
\begin{eqnarray*}
a_t(f,j)&:=&
\PP H \P{H_0}(1\otimes a(f,j))\PP{H_0}\P H\\
&=& \PP H (1\otimes a(\PP \omega f,j)) \P H.
\end{eqnarray*}
Note that 
$$H_0=A\otimes 1 +1\otimes \dg {[\omega]},$$
Assumption ${\bf (B1)}$ is as follows.
\bi
\item[(B1)] $\omega$ satisfies (1) and (2).
\bi
\item[(1)] The Lebesgue measure of $K_\omega:=\{k\in\BT|\omega(k)=0\}$ is
zero.
\item[(2)] There exists a subset $K\subset \BT$ with Lebesgue measure zero
such that
$$\omega\in C^3(\BT\setminus K)$$
and
$$ \frac{\partial\omega}{\partial k_n}(k)\not=0,\ \ \
n=1,...,d,\ \ \ k=(k_1,...,k_d)\in \BT\setminus K.$$
\eni
\eni
\begin{example}
A typical example of $\omega$ is $\omega(k)=|k|^p$ with  $p>0$.
In this case
$$K_\omega=\{0\}$$ and
$$K=\bigcup_{n=1}^d \{(k_1,...,k_d)\in\BT| k_n=0\}.$$
\end{example}

\bl{f1}
Suppose {\bf (2)} of {\bf (B1)}. Then for $f\in C_0^2(\BT\setminus K)$,
$$\left|\int_\BT e^{is\omega(k)} f(k)
dk\right|\leq \frac{c}{s^2}$$
with some constant $c$.
\el
\proof
We have,  for $1\leq m, n \leq d$,
$$e^{is\omega}=-\frac{1}{s^2}
\lk\frac{\partial\omega}{\partial k_n}\rk^{-1}
\frac{\partial}{\partial k_n}\lk
\lk\frac{\partial\omega}{\partial k_m}\rk^{-1}\frac{\partial
e^{is\omega}}{\partial k_m}\rk$$
on $\BT\setminus K$. Hence it follows that by integration by parts,
$$
\int_\BT e^{is\omega(k)} f(k)
dk
=-\frac{1}{s^2}\int_\BT
e^{is\omega(k)} \frac{\partial}{\partial k_m}\lk
\lk\frac{\partial\omega}{\partial k_m}\rk^{-1}\frac{\partial}{\partial k_n}
\lk
\lk\frac{\partial\omega}{\partial k_n}\rk^{-1}f(k)
\rk
\rk
dk.
$$
Thus we have
$$\left|
\int_\BT e^{is\omega(k)} f(k)
dk\right|
\leq \frac{1}{s^2}\int_\BT \left|
\frac{\partial}{\partial k_m}\lk
\lk\frac{\partial\omega}{\partial k_m}\rk^{-1}\frac{\partial}{\partial k_n}
\lk
\lk\frac{\partial\omega}{\partial k_n}\rk^{-1}f(k)
\rk
\rk \right|
dk.
$$
Since the integrand of the right-hand side above is integrable,  the lemma
follows.
\qed

\bp{o1}
Suppose {\bf (B1)}.
Let $f\in C^2(\BT)\cap\LT$ and $f/\sqrt\omega\in \LT$.
Then
\eq{o2}
s-\lim_{t\rightarrow \infty} a_t(f,j)\gr=0,\ \ \ j=1,...,D.
\en
\ep
\proof
Note that it follows that
$$\|a_t(f,j)\Psi\|\leq \|f/\sqrt\omega\| \|(1\otimes \dg{[\omega]}^\han)\P
H\Psi\|,
\ \ \ j=1,...,D,$$
and by the closed graph theorem,
$$\|H_0\Psi\|\leq c_1\|H\Psi\|+c_2\|\Psi\|,\ \ \ \Psi\in D(H), $$
with  some constants $c_1$ and $c_2$,
and
$$\|(1\otimes \dg{ [\omega]}^\han)\Psi\|\leq c_3\|(H_0+1)\Psi\|$$
with some constant $c_3$.
Thus it follows that
\eq{lp}
\|a_t(f,j)\Psi\|\leq  c_4 \|f/\sqrt\omega\| \|(H+1)\Psi\|
\en
with some constant  $c_4$.
Let ${\cal D}$ be a core of $A$
and
\eq{psi}
\Psi=G\otimes \add(f_1,j_1)\cdots \add(f_n,j_n)\ob,
\en
where
$G\in {\cal D}$ and $f_l\in C_0^\infty(\bk)$, $l=1,...,n$.
We see that for an arbitrary $\delta\in\RR$,
$$a(\PP{(\omega-\delta)} f,j) \Psi=\sum_{l=1}^n (\P{(\omega-\delta)}\bar f,
f_l) G\otimes
\add(f_1,j_1)\cdots \widehat{\add(f_l,j_l)} \cdots \add(f_n,j_n)\ob,$$
where $\widehat{X}$ means neglecting $X$.
Since $f f_l\in C_0^2(\bk)$,
by Lemma \ref{f1}
we see that
$$ |(\P{(\omega-\delta)}\bar f, f_l)|\leq \frac{c_5}{|t|^2} $$
with some constant $c_5$. Hence
$$s-\lim_{t\rightarrow \infty} a(\P{(\omega-\delta)} f,j) \Psi=0$$
follows.
Let ${\cal E}$ be the set of  the linear hull of vectors such as
\kak{psi}, which is  a core of $H_0$.
Thus  there exists $\Psi_\epsilon\in{\cal E}$ such that
$$\Psi_\epsilon\rightarrow \gr,\ \ \ \
H_0\Psi_\epsilon \rightarrow H_0\gr$$
strongly as $\epsilon\rightarrow 0$, which yields that
$$\lim_{\epsilon\rightarrow0}\|(H_0+1)^\han (\Psi_\epsilon-\gr)\|=0.$$
Let
$\|(H_0+1)^\han(\Psi_\epsilon-\gr)\|<\epsilon$.
 We obtain that
\begin{eqnarray*}
&&
\|a_t(f,j)\gr\|\\
&&=\|(1\otimes a(\PP{(\omega-E(H))}f,j))\gr\|\\
&&
\leq
\|(1\otimes a(\PP{(\omega-E(H))}f,j))\Psi_\epsilon\|+
\|(1\otimes a(\PP{(\omega-E(H))}f,j))(\Psi_\epsilon-\gr)\|\\
&&
\leq
\|(1\otimes a(\PP{(\omega-E(H))}f,j))\Psi_\epsilon\|+
C \|(H_0+1)^\han (\Psi_\epsilon-\gr)\|\\
&&\leq
\|(1\otimes a(\PP{(\omega-E(H))}f,j))\Psi_\epsilon\|+
C\epsilon.
\end{eqnarray*}
Then
$$\limt \|a_t(f,j)\gr\|<C\epsilon$$
for an arbitrary $\epsilon$.
Then the  proposition follows.
\qed
In addition to {\bf (B1)}, we introduce  assumptions {\bf (B2)-(B4)}.
\bi
\item[(B2)]
There exists  an operator
$$T_j(k):\FFF\rightarrow \FFF,\ \ \ k\in\BT,\ \ \ j=1,...,D,$$
such that
$$D(T_j(k))\supset D(H),\ \ \ \mbox{a. e. } k\in\BT, $$
 and
$$[1\otimes a(f,j), \hi]_W^{D(H)}(\Psi,\Phi)=\int_\BT f(k)(\Psi,
T_j(k)\Phi)dk.$$
\item[(B3)]
Let 
$\Psi \in D(H)$ and  $f\in C_0^2(\BT\setminus \widetilde K)$ with some
measurable set $\widetilde K\subset\BT$ such that $K\subset \widetilde K$
and
its Lebesgue measure is zero. Then 
$$\left|\int_{\BT} dk
f(k)(\Psi, \QQ{(H-E(H)+\omega(k))} T_j(k) \gr)\right|\in
L^1([0,\infty),ds).$$
\item[(B4)]
$\|T_j(\cdot)\gr\| \in \LT$.
\eni
\bl{key1}
Suppose  {\bf (B1)-(B4)}.
Let $f,f/\sqrt\omega\in\LT$. Then it follows that
\eq{k2}
 \int_{\BT}   \| f(k) (H-E(H)+\omega(k))^{-1}  T_j(k) \gr \| dk<\infty
\en
and
\eq{ans}
(1\otimes a(f,j))
\gr=g \int_{\BT}   f(k) (H-E(H)+\omega(k))^{-1}  T_j(k) \gr dk.
\en
\el
\proof
Noting that
$$
\|  (H-E(H)+\omega(k))^{-1}  T_j(k) \gr \|\leq \|T_j(k)\gr\|/\omega(k),\ \ \
k\not\in K_\omega,$$
we see that
\begin{eqnarray}
&& \int_{\BT}   \| f(k) (H-E(H)+\omega(k))^{-1}  T_j(k) \gr \| dk\non \\
&& \leq
\lk \int_{|k|<1}    \frac{|f(k)|^2}{\omega(k)} dk \rk^\han
\lk \int_{|k|<1}
\omega(k) \|  (H-E(H)+\omega(k))^{-1}  T_j(k) \gr \| ^2 dk\rk^\han \non \\
&&+
\lk \int_{|k|\geq 1}    |f(k)|^2 dk \rk^\han
\lk \int_{|k|\geq 1} \|(H-E(H)+\omega(k))^{-1}  T_j(k) \gr \| ^2 dk\rk^\han
\non\\
&&\leq (\|f/\sqrt\omega\|+\|f\|)\|T_j(\cdot)\gr\|<\infty.
\end{eqnarray}
Then \kak{k2} follows.
We divide a proof of \kak {ans} into three  steps. \\
{\bf (Step 1)} Let $f\in C_0^2(\BT\setminus\widetilde{K})$,
$f/\sqrt\omega\in \LT$,
and $\Psi,\Phi\in D(H)$.
Then
\eq{sa1}
(\Psi, (1\otimes a( f,j))\gr)=
-ig
\int_0^\infty
\lk
\int_{\BT} (\Psi, f(k) \QQ{(H-E(H)+\omega(k))} T_j(k) \gr) dk\rk ds.
\en
{\it Proof of Step 1}\\
Let $\Psi,\Phi\in {\cal D}:=C_0^\infty(\BT)\otimes
 D(\dg{[\omega]})$.
Note that ${\cal D}$ is a core of $H$.
We see that by  \kak{to1} of Proposition \ref{well} and {\bf (B2)},
\begin{eqnarray*}
&&\frac{d}{dt}(\Psi, a_t(f,j)\Phi)\\
&&
=-i(H\P H \Psi, (1\otimes a(\PP\omega f))\P H\Phi)-i(\P H \Psi, (1\otimes
a(\omega\PP\omega f))\P H\Phi)\\
&& \hspace{4cm}+i((1\otimes \add(\P\omega \bar f)) \P H\Psi, H\P H\Phi)\\
&&=-ig(\hi\P H \Psi, (1\otimes a(\PP\omega f))\P H\Phi)+
ig((1\otimes \add(\P\omega \bar f))\P H\Psi, \hi \P H\Phi)\\
&&=ig[1\otimes a(\PP\omega f),\hi]_W^{D(H)}(\P H\Psi, \P H\Phi)\\
&&=ig\int_{\BT} f(k) \PP{\omega(k)}(\Psi, \PP H T_j(k) \P H\Phi) dk.
\end{eqnarray*}
Then we obtain that for $\Psi,\Phi\in {\cal D}$,
\begin{eqnarray}
&& \hspace{-1cm}
(\Psi, a_t(f,j)\Phi)\nonumber \\
&& \label{psi2}
\hspace{-1cm}=(\Psi, (1\otimes a(f,j))\Phi)+
ig\int_0^t
\lk
\int_{\BT} f(k) \QQ{\omega(k)}(\Psi, \QQ H T_j(k) \Q H\Phi) dk\rk ds.\ \ \ \
\
 \ \
\end{eqnarray}
Let $\Psi,\Phi\in D(H)$.
There exist  sequences $\Psi_m, \Phi_n\in {\cal D}$ such that
$\limm \Psi_m=\Psi$ and $\limn \Phi_n=\Phi$ strongly.
\kak{psi2} holds true for $\Psi,\Phi$ replaced by $\Psi_m, \Phi_n$,
respectively.
By a simple limiting argument as $m\rightarrow \infty$ and then
$n\rightarrow \infty$,
we get \kak{psi2} for $\Psi,\Phi\in D(H)$.
By Proposition \ref{o1} and \kak{psi2} we have
\begin{eqnarray*}
&&0=\lim_{t\rightarrow \infty}(\Psi,a_t(f,j)\gr)\\
&&=(\Psi, (1\otimes a( f,j))\gr)+
ig
\int_0^\infty
\lk
\int_{\BT} (\Psi, f(k) \QQ{(H-E(H)+\omega(k))} T_j(k) \gr) dk\rk ds.
\end{eqnarray*}
Thus \kak{sa1} follows.
\qed
\noindent
{\bf (Step 2)} 
\kak{ans} holds true for $f$ such that 
$f\in C_0^2(\BT\setminus\widetilde{K})$ and
$f/\sqrt\omega\in \LT$.

{\it Proof of Step 2}\\
By {\bf (B3)} and the Lebesgue dominated convergence theorem, we have
\begin{eqnarray*}
&& -ig
\int_0^\infty
\lk
\int_{\BT} (\Psi, f(k) \QQ{(H-E(H)+\omega(k))} T_j(k) \gr) dk\rk ds \\
&&
=-ig\lim_{\epsilon\rightarrow 0}
\int_0^\infty ds e^{-\epsilon s}
\lk
\int_{\BT} (\Psi, f(k) \QQ{(H-E(H)+\omega(k))} T_j(k) \gr) dk\rk.
\end{eqnarray*}
By {\bf (B4)},
\begin{eqnarray*}
&&
\int_{\BT} dk \int_0^\infty
\left|
e^{-s\epsilon}
(\Psi, f(k) \QQ{(H-E(H)+\omega(k))}  T_j(k) \gr)
\right|ds \\
&&\leq  \|\Psi\|
\lk
\int_{\BT}
|f(k)| \|T_j(k) \gr\| dk\rk
\int_0^\infty
e^{-s\epsilon}ds
<\infty.
\end{eqnarray*}
Hence Fubini's theorem yields that $\int dk$ and $\int ds$ can be exchanged,
i.e.,
\begin{eqnarray*}
&&
-ig\lim_{\epsilon\rightarrow 0}
\int_0^\infty
e^{-\epsilon s}
\lk
\int_{\BT} (\Psi, f(k) \QQ{(H-E(H)+\omega(k))} T_j(k) \gr) dk\rk ds \\
&&
=-ig\lim_{\epsilon\rightarrow 0}
\int_{\BT}
\lk
\int_0^\infty
(\Psi, f(k) \QQ{(H-E(H)+\omega(k)-i\epsilon)} T_j(k) \gr) ds \rk dk \\
&&=
g\lim_{\epsilon\rightarrow 0}
\int_{\BT}
(\Psi, f(k) (H-E(H)+\omega(k)-i\epsilon)^{-1}  T_j(k) \gr) dk.
\end{eqnarray*}
We can check that, for $k\not\in K_\omega$,
\begin{eqnarray}
&&
|(\Psi, f(k) (H-E(H)+\omega(k)-i\epsilon)^{-1}  T_j(k) \gr)|\non \\
&&
\label{ppp}
\hspace{2cm}
\leq
\|\Psi\| |f(k)| \| (H-E(H)+\omega(k))^{-1}  T_j(k) \gr\|,\\
&&
 \int_{\BT}
|f(k)| \| (H-E(H)+\omega(k))^{-1}  T_j(k) \gr\| dk \non\\
&& \label{b4} \hspace{2cm} \leq
(\|f/\sqrt\omega\|+\|f\|)\|T_j(\cdot)\gr\|<\infty,\\
&&
\label {pp}
 s-\lim_{\epsilon\rightarrow 0} (H-E(H)+\omega(k)-i\epsilon)^{-1}\gr=
(H-E(H)+\omega(k))^{-1}\gr,
\end{eqnarray}
\kak{ppp}, \kak{b4} and \kak{pp} imply
 that by the Lebesgue dominated convergence theorem,
\begin{eqnarray*}
&& g\lim_{\epsilon\rightarrow 0}
\int_{\BT}
(\Psi, f(k) (H-E(H)+\omega(k)-i\epsilon)^{-1}  T_j(k) \gr) dk \\
&&=
g
\int_{\BT}
(\Psi, f(k) (H-E(H)+\omega(k))^{-1}  T_j(k) \gr) dk.
\end{eqnarray*}
Since, by \kak{b4}
we have
\begin{eqnarray*}
(\Psi, a(f,j)\gr)
&=&
g
\int_{\BT}
(\Psi, f(k) (H-E(H)+\omega(k))^{-1}  T_j(k) \gr) dk\\
&=&
(\Psi, g \int_{\BT}   f(k) (H-E(H)+\omega(k))^{-1}  T_j(k) \gr dk ), 
\end{eqnarray*}
we obtain \kak{ans}. \qed
{\bf (Step 3)} 
\kak{ans} holds true for  $f$ such that 
$f,f/\sqrt\omega\in \LT$. 

{\it Proof of Step 3}\\
 Set
$$g(k):=\lkk\begin{array}{ll} f(k)/\sqrt{\omega(k)},&|k|<1,\\
f(k),&|k|\geq 1.
\end{array}
\right.
$$
Since $g\in\LT$, there exists a sequence $g_\epsilon\in
C_0^\infty(\BT\setminus\widetilde K)$ such that
$g_\epsilon\rightarrow g$ strongly as $\epsilon\rightarrow 0$.
Define
$$f_\epsilon(k) :=\lkk\begin{array}{ll} \sqrt{\omega(k)} g_\epsilon(k)
,&|k|<1,\\
g_\epsilon(k),&|k|\geq 1.
\end{array}
\right.
$$
Hence $f_\epsilon\in C_0^3(\BT\setminus \widetilde K)$ by (2) of {\bf (B1)},
and
it follows that
\begin{eqnarray}
&&
\label{s1}
\int_\BT
\left|f(k)-f_\epsilon(k)\right|^2/\omega(k) dk\rightarrow 0,\\
&&
\label{s2}
\int_{|k|>1}
\left|f(k)-f_\epsilon(k)\right|^2dk\rightarrow 0,
\end{eqnarray}
as $\epsilon\rightarrow 0$.
We see that, by \kak{s1} and  \kak{s2},
$$\|(1\otimes a(f))\gr-(1\otimes a(f_\epsilon))\gr\|\leq
\|(f-f_\epsilon)/\sqrt\omega\| \|(1\otimes
\dg{[\omega]}^\han )\gr\|\rightarrow 0$$
and
\begin{eqnarray*}
&&
\left\|  \int_{\BT}  ( f(k)-f_\epsilon(k))
(H-E(H)+\omega(k))^{-1}  T_j(k) \gr dk\right\|\\
&&
\leq \lkk
\lk \int_{|k|<1} \frac{|f(k)-f_\epsilon(k)|^2}{\omega(k)}dk\rk^\han +
\lk \int_{|k|\geq 1} |f(k)-f_\epsilon(k)|^2dk\rk^\han\rkk
\|T_j(\cdot)\gr\|\\
&& \rightarrow 0
\end{eqnarray*}
as $\epsilon\rightarrow 0$.
Then we
can extend \kak{ans} to $f$ such that
$f,f/\sqrt\omega\in\LT$.
\qed
\subsection{Main theorem I}
\label{w3}
Set
$$\k(k):=(H-E(H)+\omega(k))^{-1}  T_j(k) \gr,\ \ \ k\not\in K_\omega.$$
We define  $$\T_j:\LT\rightarrow \FFF,\ \ \ \ j=1,...,D,$$  by
$$\T_j f:=\int_\BT f(k)\k(k) dk,$$
where the integral is taken in the strong sense in $\FFF$.
I.e., we have
\eq{ttt}
(1\otimes a(f,j))\gr=g\T_j f.
\en
\bp{Carl}
(1) $\T_j$ is a Hilbert-Schmidt operator if and only if
$$\int_\BT \|\k(k)\|^2 dk<\infty.$$
(2) Suppose that $\T_j$ is a Hilbert-Schmidt operator,
Then
$$\sum_{m=1}^\infty \|\T_j e_m\|^2= \int_\BT \|\k(k)\|^2 dk$$
for an arbitrary complete orthonormal system $\{e_m\}_{m=1}^\infty$ in
$\LT$.
\ep
\proof
The adjoint of $\T_j$,
$$\TT_j :\FFF\rightarrow \LT,\ \ \ j=1,...,D,$$
is referred to as
a Carleman operator   (see e.g., \cite[p.141]{wei})
with kernel $\k$, i.e.,
$$\TT_j \Phi(\cdot):=(\k(\cdot), \Phi).$$
It is known \cite[Theorem 6.12]{wei}
that $\TT_j$ is a Hilbert-Schmidt operator if and only if
$\int_\BT \|\k(k)\|^2 dk<\infty$.
Moreover suppose that $\TT_j$ is a Hilbert-Schmidt operator.
Then
$$\tr(\T_j\TT_j)=\int_\BT \|\k(k)\|^2 dk $$ is also known,
which implies that
$\T_j$ is a Hilbert-Schmidt operator if and only if
$\int_\BT \|\k(k)\|^2 dk<\infty$, and
$$\tr (\TT_j \T_j)=\tr(\T_j\TT_j)=\int_\BT \|\k(k)\|^2 dk .$$
Thus the proposition follows.
\qed

The main theorem in this section is as follows.
\bt{asym}
Suppose  {\bf (B1)-(B4)}.
Then (1), (2) and (3) are equivalent.
\bi
\item[(1)]
$P_H\FFF\subset D(1\otimes N^\han)$,
\item[(2)]
$\T_j$  is a Hilbert-Schmidt operator for all $j=1,...,D$ and all
$\gr\in\pg\FFF$,
\item[(3)]
$ \int_{\BT}\|(H-E(H)+\omega(k))^{-1}T_j(k)\gr\|^2 dk<\infty$
for all $j=1,...,D$ and all $\gr\in\pg\FFF$.
\eni
Suppose that one of (1), (2) and  (3) holds,
it follows that for an arbitrary ground state $\gr$,
\eq{mai}
\|(1\otimes N^\han)\gr\|^2= g^2\sum_{j=1}^D
\int_{\BT}\|(H-E(H)+\omega(k))^{-1}T_j(k)\gr\|^2 dk.
\en
\et
\proof
Let $\{e_m\}_{m=1}^\infty$ be a complete orthonormal system of $\LT$ such
that $e_m/\sqrt\omega\in\LT$.
It is proven in Lemma \ref{g4} that
$\pg\FFF\subset D(1\otimes N^\han)$ if and only if
\eq{sa3}
\sum_{j=1}^D \sum_{m=1}^\infty \|(1\otimes a(\ov{e_m},j))\gr\|^2<\infty
\en
for an arbitrary $\gr\in \pg\FFF$.
By \kak{ttt},
$$(1\otimes  a(\ov{e_m},j))\gr=g\T_j \ov{e_m}.$$
Hence
$\pg\FFF\subset D(1\otimes N^\han)$ if and only if
$$g^2 \sum_{j=1}^D\sum_{m=1}^\infty \|\T_j \ov{e_m}\|^2<\infty,\ \ \ \gr\in\pg\FFF.$$
That is to say,
$\pg\FFF\subset D(1\otimes N^\han)$ if and only if
$\T_j$ is a Hilbert-Schmidt operator for all $j=1,...,D$, and all $\gr\in \pg\FFF$, 
i.e.,
by Proposition \ref{Carl}, $\pg\FFF\subset D(1\otimes N^\han)$ if and only
if
$$g^2 \sum_{j=1}^D\int_\BT\|\kap_j(k)\|^2 dk<\infty,\ \ \ \gr\in\pg\FFF.$$
Then the first half of the theorem is proven.
 Moreover by Lemma \ref{g4}, when $\gr\in D(1\otimes N^\han)$,
$$\|(1\otimes N^\han)\gr\|^2=
\sum_{j=1}^D \sum_{m=1}^\infty \|(1\otimes a(\ov{e_m},j))\gr\|^2,$$
which yields that
$$\|(1\otimes N^\han)\gr\|^2=g^2 \sum_{j=1}^D \tr (\TT_j \T_j)= g^2
\sum_{j=1}^D \int_\BT \|\kap_j(k)\|^2 dk.$$
Thus the proof is complete.
\qed
\begin{remark}
In \cite{ahh2} a more general formula than \kak{mai} is obtained.
\end{remark}

\section{Proof of $\ma(H)\leq \ma(A)$ }
\subsection{Quadratic forms}
We revive $H=H_0+g\hi$, where $H_0=A\otimes 1+1\otimes\dg S$, 
 and ${\bf (A.1)}-{\bf (A.3)}$ are assumed. 
Set
$$\bh:=H_0-E(H_0).$$
Actually
$$E(H_0)=E(A).$$
The quadratic form $\beta_0$ associated with $\bh$ is defined by 
$$\beta_0(\Psi, \Phi):=(\bh^\han \Psi, \bh^\han \Phi),\ \ \ \Psi,\Phi\in
D(\bh^\han).$$
Define a symmetric form by 
$$\beta_\hi(\Psi,\Phi):=(\Psi, {\hi}\Phi),\ \ \ \Psi,\Phi\in D(\bh).$$
Since $\|\hi \Psi\|\leq a\|H_0\Psi\|+b\|\Psi\|$, 
it follows that 
$$\|\hi \Psi\|\leq a\|\bh\Psi\|+b'\|\Psi\|,$$
where 
$b'=b+ a |E(H_0)|$. 
Then 
$\hi(\bh+\mu)\f$ and $(\bh+\mu)\f\hi$, $\mu>0$,  are bounded operators
with
$$\|\hi(\bh+\mu)\f\|\leq a+b'/\mu,\ \ \ \|(\bh+\mu)\f\hi\|\leq a+b'/\mu.$$ 
By an interpolation argument \cite[Section IX]{rs2},
$(\bh+\mu)^{-\han}\hi(\bh+\mu)^{-\han}$ is also a bounded operator with
$$\| (\bh+\mu)^{-\han}\hi(\bh+\mu)^{-\han}\|\leq a+b'/\mu.$$
Then
\eq{plm}
|\beta_\hi(\Psi,\Psi)|\leq (a+b'/\mu) \beta_0(\Psi,\Psi)+ (a+b'/\mu)\|\Psi\|^2,\ \ \ 
\Psi\in D(\bh), 
\en
for an arbitrary $\mu>0$.
By \kak{plm}, a polarization identity \cite{rs1} and a limiting argument,
$\beta_\hi(\Psi, \Phi)$ can be
extended  to $\Psi, \Phi\in D(\bh^\han)$. 
The extension of $\beta_\hi$ is denoted by $\widetilde{\beta}_\hi$,  and  which satisfies 
\eq{plmm}
|\widetilde{\beta}_\hi(\Psi,\Psi)|\leq (a+b'/\mu) \beta_0(\Psi,\Psi)+ (a+b'/\mu)\|\Psi\|^2,\ \ \ 
\Psi\in D(\bh^\han).
\en
Thus
we see that, for a sufficiently small $g$, 
$$\beta_H:=\beta_0+g\widetilde{\beta}_\hi$$
is a semibounded closed quadratic form 
on $D(\bh^\han)\times D(\bh^\han)$. 
Then  by the representation theorem for forms 
\cite[p.322, Theorem 2.1]{kato}, 
 there exists a unique  self-adjoint operator $H'$  such that 
$D(H')\subset D(\bh^\han)$ and 
$$\beta_H(\Psi,\Phi)=(\Psi, H'\Phi),\ \ \ 
\Psi\in D(\bh^\han), \ \ \ \Phi\in D(H').$$
On the other hand,  we can see directly that 
$D(H)\subset D(\bh^\han)$ and 
$$\beta_H(\Psi, \Phi)=(\Psi, H\Phi),\ \ \ \Psi\in D(\bh^\han), \ \ \ \Phi\in D(\bh).$$
which yields that 
$$H'=H.$$
I.e., $H$ is a unique self-adjoint operator associated with the quadratic form $\beta_H$.
We generalize this fact in the next subsection. 

\subsection{Abstract results}
As was seen in the previous subsection, self-adjoint operator $H=H_0+g\hi$ is defined through 
the quadratic form $\beta_H$. 
In this subsection, as a mathematical generalization,  
we define a total Hamiltonian $\hq$ 
through an abstarct quadratic form, and estimate an upper bound of 
${\rm dim} \lkk P_\hq\FFF\cap D(1\otimes N^\han)\rkk $. 
\begin{remark}
Hamiltonians of the Nelson model {\rm without} ultraviolet cutoffs are 
defined as the self-adjoint operator associated with a semibounded quadratic form. 
See \cite{hhs, ne3}. As far as we know, it can not be represented as the form $H_0+g\hi$. 
\end{remark}

Let $\betai$
be a symmetric quadratic form with  form  domain
$D(\bh^\han)$
such that
\eq{21}
|\betai(\Psi,\Psi)|\leq a\beta_0(\Psi,\Psi)+b(\Psi,\Psi),\ \ \ \ \Psi\in
D(\bh^\han),
\en
with some nonnegative constants $a$ and $b$.
Define the quadratic form $\beta$ on $D(\bh^\han)$ by
$$\beta:=\beta_0+g \betai.$$
\begin{proposition}
\label{ha}
Let $|g |< 1/a$. Then there exists a unique self-adjoint operator
$\hq$ associated with $\beta$ such that
its  form domain is
$D(\bh^\han)$,
$$\beta(\Psi,\Phi)=(\Psi, \hq \Phi),\ \ \ \ \Psi\in D(\bh^\han),\Phi\in
D(\hq ),$$
and
$$\beta(\Psi,\Phi)=(\hq _+^\han \Psi, \hq _+^\han \Phi)-(\hq _-^\han\Psi,
\hq _-^\han
\Phi),
\ \ \ \ \Psi,\Phi\in D(\bh^\han),$$
where
$$\hq _+:=\hq E_\hq ((0,\infty)),\ \ \ \ \hq _-:=-\hq E_\hq ((-\infty,0]).
$$
\end{proposition}
\proof
From  \kak{21} it follows that
$$|g\betai(\Psi,\Psi)|\leq |g|a\beta_0(\Psi,\Psi)+|g|b(\Psi,\Psi).$$
Hence by the KLMN theorem \cite[Theorem X.17]{rs2}, the  proposition
follows.
\qed

Assumptions {\bf (Gap)} and {\bf (N)} are  as follows.
\bi
\item[(Gap)] $\is_{\rm ess} (A)-E(A)>0$.
\item[(N)]
$\displaystyle
\lim_{g\rightarrow 0}
\sup_{\Psi\in (\pgq\FFF) \cap D(1\otimes N^\han)}
\frac{\|(1\otimes N^\han)\Psi\|}{\|\Psi\|} =0
$.
\eni
Suppose that
$\sigma_{\rm p}(S)\not\ni 0$.
Then by the facts that
\begin{eqnarray*}
&& \is (\dg S\lceil_{\oplus_{n=1}^\infty [\otimes_s^n{\cal W}]})\geq 0,\\
&& \sigma_{\rm p} (\dg S\lceil_{\oplus_{n=1}^\infty [\otimes_s^n{\cal
W}]})\not\ni 0,\\
&& \sigma(\dg S\lceil_{\otimes_s^0{\cal W}})=\sigma_{\rm p}
(\dg S\lceil_{\otimes_s^0{\cal W}})=\{0\},
\end{eqnarray*}
$\dg S$ is nonnegative self-adjoint operator, and has a unique ground state
${\obb}$ with eigenvalue $0$.
We have a lemma.
\bl{1}
Assume {\bf (A1)},   {\bf (A2)}, {\bf (Gap)}, {\bf (N)} and
$\sigma_{\rm p}(S)\not\ni 0$.
Then there exists $\delta(g)>0$ such that
$$\lim_{g\rightarrow 0}\delta(g)=0$$
and,
for $g$ with $\delta (g)<1$,
$${\rm dim} \lkk (\pgq\FFF)\cap D(1\otimes N^\han)\rkk
\leq \frac{1}{1-\delta(g)} \ma(A).$$
\el
\proof
Let $\epsilon>0$ be such that
$$[E(A), E(A)+\epsilon)\cap \sigma(A)=\{E(A)\}$$
and we set
$$\qqq:=E_A([E(A), E(A)+\epsilon)),\ \ \ \
\qqq^\perp:=1-\qqq.$$
Furthermore let
$$\qq:=E_{\dg S}(\{0\}).$$
We fix a  $\gr\in (\pgq\FFF)\cap D(1\otimes N^\han)$.
Using the inequality
$$1\otimes 1\leq 1\otimes N+1\otimes \qq$$
in the sense of form,
we have
\begin{eqnarray}
(\gr, \gr)
&\leq& ((1\otimes N^\han)\gr,(1\otimes N^\han)\gr)
+(\gr,(1\otimes \qq)\gr)\non \\
\label{2}
&\leq& \|(1\otimes N^\han )\gr\|^2
+\|(\qqq\otimes \qq )\gr\|^2+\|(\qqq^\perp\otimes \qq) \gr\|^2.
\end{eqnarray}
Let $Q:=\qqq^\perp\otimes \qq$.
In Proposition \ref{p} we shall show that
\eq{29}
\gr\in D(\bh^\han),\ \ \
 Q\gr\in D(\bh^\han)\en
and
\eq{31}
\bh^\han Q\gr=Q\bh^\han \gr.
\en
Hence we have
\begin{eqnarray*}
0&=& (Q\gr, (\hq -E(\hq ))\gr)\\
&=&\beta(Q\gr,\gr)-E(\hq )(Q\gr,\gr)\\
&=&\beta_0(Q\gr,\gr)+g \betai(Q\gr,\gr)-E(\hq )(Q\gr,\gr).
\end{eqnarray*}
From this we have
\eq{35}
-g \betai(Q\gr,\gr)=
(\bh^\han Q\gr,\bh^\han \gr)-E(\hq )(Q\gr,\gr).
\en
Since we have by \kak{31}
\begin{eqnarray*}
&&
(\bh^\han Q\gr,\bh^\han\gr)\\
&&=(\bh^\han Q\gr,\bh^\han Q\gr)\\
&&=
\int_{[{E(A)},\infty)\times[0,\infty)}(\lambda+\mu-E(A))d\|
(E_A(\lambda)\otimes E_{\dg S}(\mu)) Q\gr\|^2 \\
&&=
\int_{[{E(A)}+\epsilon, \infty)\times\{0\}}(\lambda+\mu-E(A))
d\| (E_A(\lambda)\otimes E_{\dg S}(\mu)) Q\gr\|^2 \\
&& \geq \epsilon(\gr, Q\gr),
\end{eqnarray*}
then \kak{35} implies that
\eq{32}
-g \betai(Q\gr,\gr)
\geq (\epsilon-E(\hq ))(Q\gr,\gr).
\en
We  shall estimate $|\betai(Q\gr, \gr)|$.
\begin{eqnarray*}
\beta_0(\gr,\gr)
&=&(\gr, \hq \gr)-g\betai(\gr,\gr)\\
&\leq& E(\hq )\|\gr\|^2+|g|\lk a\beta_0(\gr,\gr)+b(\gr.\gr)\rk,
\end{eqnarray*}
which yields that, since $|g|<1/a$,
$$\beta_0(\gr,\gr)\leq \frac{E(\hq )+|g|b}{1-a|g|}(\gr,\gr).$$
Then we have
$$|\betai(\gr,\gr)|\leq
\lk a\beta_0(\gr,\gr)+b(\gr,\gr)\rk\leq
\ci (\gr,\gr),$$
where
$$\ci:=\frac{a(E(\hq )+|g|b)}{1-a|g|}+b.$$
From  the polarization identity
\begin{eqnarray*}
&& \betai(Q\gr,\gr)=  \frac{1}{4}
\lkk\lk\betai((1+Q)\gr,(1+Q)\gr)-\betai((1-Q)\gr,(1-Q)\gr)\rk\right.\\
&&\hspace{2cm}\left.-i\lk\betai((1+iQ)\gr,(1+iQ)\gr)-\betai((1-iQ)\gr,(1-iQ)
\gr)\rk\rkk,
\end{eqnarray*}
it follows that
\eq{23}
|\betai(Q\gr,\gr)|
\leq
2\ci (\gr,\gr).
\en
Note that
$$|\beta(\Psi,\Psi)-\beta_0(\Psi,\Psi)|=|g||\betai(\Psi,\Psi)|\leq
|g|(a+b)\|(\bh+1)^\han\Psi\|^2.$$
Then
$$\lim_{g\rightarrow 0}
\sup_{\Psi\in D(\bh^\han)}
\frac{|\beta(\Psi,\Psi)-\beta_0(\Psi,\Psi)|}{\|(\bh+1)^\han\Psi\|^2}\leq
\lim_{g\rightarrow 0}|g|(a+b)=0,$$
which implies that
for $z\in\CC$ with $\Im z\not=0$,
\eq{chi}
\lim_{g\rightarrow 0}
\|(\hq -z)\f-(\bh-z)\f\|=0.
\en
See Proposition \ref{ko-taro-} for a proof of \kak{chi}.
Thus it follows that
\eq{24}
\lim_{g\rightarrow 0} E(\hq )=E(\bh)=0.
\en
Then there exists a constant $c>0$ such that
for all $g$ with $|g|<c$,
it obeys that
$$\epsilon-E(\hq )>0.$$
Then by \kak{32} and  \kak{23}, for $g$ with $|g|<c$,
$$\|Q\gr\|^2\leq
|g|\frac{|\betai(Q\gr,\gr)|}{\epsilon-E(\hq )}
\leq
2|g|\frac{\ci}
{\epsilon -E(\hq)}\|\gr\|^2.$$
Let
$$c(g):=\sup_{\Psi\in (\pgq\FFF) \cap D(1\otimes N^\han)}
\frac{\|(1\otimes N^\han)\Psi\|}{\|\Psi\|}.$$
Together with \kak{2} we have
\eq{311}
(\gr, \gr)\leq
c(g)^2\|\gr\|^2+
2|g|\frac{\ci}
{\epsilon -E(\hq )}\|\gr\|^2
+\|(\qqq\otimes \qq)\gr\|^2.
\en
Setting
$$\delta(g):=
c(g)^2+
2|g|\frac{\ci}
{\epsilon -E(\hq )},
$$
we see that by \kak{24} and {\bf (N)},
$$\lim_{g\rightarrow 0}\delta(g)=0.$$
Then by \kak{311} there exists $g_\ast\leq  c$ such that for $g$ with
$|g|<g_\ast$,
\eq{111}
(\gr, \gr)\leq
(1-\delta(g))\f
(\gr, (\qqq\otimes \qq) \gr).
\en
Let $\{\gr^j\}_{j=1}^M$, $M\leq\infty$, be a complete orthonormal system of
$(\pgq\FFF)\cap D(1\otimes N^\han)$.
Then by \kak{111},
\eq{1111}
(\gr^j, \gr^j)\leq
(1-\delta(g))\f
(\gr^j, (\qqq\otimes \qq) \gr^j).
\en
Summing up from $j=1$ to $M$, we have
$$
\dr \lkk (\pgq\FFF)\cap D(1\otimes N^\han)\rkk
\leq
(1-\delta(g))\f
\sum_{j=1}^M  (\gr^j, (\qqq\otimes \qq)\gr^j).$$
Since
\begin{eqnarray*}
\sum_{j=1}^M  (\gr^j, (\qqq\otimes \qq)\gr^j)
&=&
\sum_{j=1}^M  (\gr^j, (P_A\otimes \qq)\gr^j)\\
& \leq&  \tr (P_A\otimes \qq)\\
&=& \tr P_A \times \tr \qq\\
&=& \ma(A),
\end{eqnarray*}
we obtain that
$$\dr \lkk (\pgq\FFF)\cap D(1\otimes N^\han)\rkk  \leq (1-\delta(g))\f
\ma(A).$$
Thus the lemma is proven.
\qed
From Lemma \ref{1},
corollaries immediately follow.
\bc{9}
We assume the same assumptions as in Lemma \ref{1}.
Suppose that
\eq{dn}
\pgq\FFF\subset D(1\otimes N^\han).
\en
Then
$$\ma(\hq) \leq (1-\delta(g))\f \ma(A).$$
\ec
In addition, suppose that  $g$ is such that $\delta(g)<1/2$ and $\ma(A)=1$.
Then
$\ma(H)=1$.
\proof
Since $\pgq\FFF\cap D(1\otimes N^\han)=\pgq\FFF$,
the corollary follows from Lemma \ref{1}.
\qed
\bc{4}{\bf [Overlap]}
We assume   the same assumptions as in Lemma \ref{1} and  \kak{dn}. 
Let $g$ be  such that $\delta (g)<1$.
Then for an arbitrary ground state $\gr$, it follows that
$$(\gr, (\pa\otimes \qq) \gr)\not=0.$$
\ec
\proof
By \kak{111} it is seen that
$$0<\|\gr\|^2\leq(1-\delta(g))\f(\gr, (\qqq\otimes \qq)\gr)=
(1-\delta(g))\f(\gr, (\pa\otimes \qq)\gr).$$
Hence  the corollary follows.
\qed

\subsection{Main theorem II}
We assume \kak{w5} and \kak{w6}, i.e., $H=H_0+g\hi$ and 
$H_0=A\otimes 1+1\otimes \dg{[\omega]}$.  
Now we are in the position to state the main theorem in this section. 
\bt{main2}
Supose that {\bf (B1)}-{\bf (B4)}, {\bf (A1)}, {\bf (A3)}, {\bf (Gap)}.
We assume that
$$\int_{\BT} \|(H-E(H)+\omega(k))^{-1}T_j(k)\gr\|^2 dk <\infty, $$
and
\eq{mai2}
\sup_{\gr\in P_H\FFF} \frac{\sum_{j=1}^D
\int_{\BT}\|(H-E(H)+\omega(k))^{-1}T_j(k)\gr\|^2}{\|\gr\|^2} dk <\infty.
\en
Then there exists a constant $g_\ast$ such that
for $g$ with $|g|<g_\ast$, it follows that
$$\ma(H)\leq \ma(A).$$
\et
\proof
By Theorem \ref{asym}, it follows that $P_H\FFF\subset D(1\otimes N^\han)$
and
$$\|(1\otimes N^\han)\gr\|^2=
g^2 \sum_{j=1}^D
\int_{\BT}\|(H-E(H)+\omega(k))^{-1}T_j(k)\gr\|^2 dk.$$
By  \kak{mai2} we  have
\begin{eqnarray*}
&& \lim_{g\rightarrow 0}
\sup_{\Psi\in P_H\FFF}
\frac{\|(1\otimes N^\han)\gr\|}{\|\gr\|}\\
&&
=
\lim_{g\rightarrow 0} |g| \sup_{\gr\in P_H\FFF}
\lk
\frac{\sum_{j=1}^D
\int_{\BT}\|(H-E(H)+\omega(k))^{-1}T_j(k)\gr\|^2 dk}{\|\gr\|^2}\rk^\han
=0.
\end{eqnarray*}
From this  and  Corollary  \ref{9}, the theorem follows.
\qed

\section{Examples}
\subsection{GSB models}
GSB models are  a generalization of the spin-boson model,
which was introduced and investigated in \cite{ah1}.
Examples of GSB models are e.g.,
$N$-level systems coupled to a Bose field, lattice spin systems,
the Pauli-Fierz model with the dipole approximation neglected $A^2$ term,
a Fermi field coupled to a Bose field, etc.
See \cite [p. 457]{ah1}.

The Hilbert space on which  GSB Hamiltonians act is
$$\FFFg:=\hhh\otimes\fff(\LT),$$
where $\hhh$ is a Hilbert space.
Let $a(f)$
and $\add(f),f\in\LT$,
be the annihilation operator and the creation operator on $\fff(\LT)$,
respectively. We use the same notations $a(f)$ and $\add(f)$ as those of
Subsection \ref{s12}.
We set
$$\phi(\lambda):=\frac{1}{\sqrt2}(\add(\bar \lambda)+a(\lambda)),\ \ \
\lambda\in\LT.$$
GSB Hamiltonians are defined by
$$\gsb:=\gsbb+\alpha \gsbbb.$$
Here $\alpha \in\RR$ is a coupling constant, and
\begin{eqnarray*}
&& \gsbb:=A\otimes 1+1\otimes \dg \ogsb ,\\
&&\gsbbb:=\ov{\sum_{j=1}^J B_j\otimes \phi(\lambda_j)},
\end{eqnarray*}
where
$\ogsb :  \LT\rightarrow \LT$ is a multiplication operator
by $\ogsb (k)$ such that
$$\ogsb (\cdot):\BT\rightarrow [0,\infty)$$ and
$\ov{X}$ denotes the closure of $X$.
Assumption {\bf (GSB1)}-{\bf (GSB5)} are as follows.
\bi
\item[(GSB1)] Operator $A$ satisfies {\bf (A1)}. Set $\ba:=A-E(A)$.
\item[(GSB2)] $\lambda_j,\lambda_j/\sqrt\ogsb \in\LT$, $j=1,...,J$.
\item[(GSB3)] $B_j$, $j=1,...,J$,  is a symmetric operator,
$D(\ba^\han)\subset \cap_{j=1}^JD(B_j)$ and
there exist constants $a_j$ and $b_j$ such that
$$\|B_jf\|\leq a_j\|\ba^\han f\|+b_j\|f\|,\ \ \ f\in D(\ba^\han).$$
Moreover
$$|\alpha|<\lk\sum_{j=1}^Ja_j\|\lambda_j/\sqrt\ogsb \|\rk^{-1}.$$
\item[(GSB4)] $\ogsb $ satisfies that
\bi
\item[(1)] $\ogsb (\cdot)$ is continuous,
\item[(2)] $\lim_{|k|\rightarrow \infty}\ogsb (k)=\infty$,
\item[(3)]
there exist constants $C>0$ and $\gamma>0$ such that
$$|\ogsb (k)-\ogsb (k')|\leq C|k-k'|^\gamma(1+\ogsb (k)+\ogsb (k')).$$
\eni
\item[(GSB5)] $\lambda_j$, $j=1,...,J$,  is continuous.
\eni
\bp{sb1}
Assume {\bf (GSB1)-(GSB3)}.
Then $\gsb$ is self-adjoint on
$$D(\gsbb)=D(A\otimes 1)\cap D(1\otimes \dg \ogsb )$$ and bounded from
below.
Moreover it is essentially self-adjoint on any core of $\gsbb$.
\ep
\proof
We can show that
\eq{m1}
\|\gsbbb \Psi\|\leq
\lk\sum_{j=1}^Ja_j\|\lambda_j/\sqrt\ogsb \|\rk\|\gsbb\Psi\|+b\|\Psi\|
\en
for $\Psi\in D(\gsbb)$ with some constant $b$.
Then by the Kato-Rellich theorem, the proposition follows.
for details. \qed
We introduce  assumptions.
\bi
\item[(IR)] $\lambda_j/\ogsb \in\LT$, $j=1,...,J$.
\item[(GSB6)] $\ogsb $ satisfies {(\bf B1)} with $\omega$ replaced by
$\ogsb$.
\item[(GSB7)] $\lambda_j\in C^2(\BT)$, $j=1,...,J$.
\eni

\bp{sb2}
We assume  {\bf (GSB1)-(GSB5)}, {\bf (IR)} and {\bf (Gap)}.
Then there exists a constant $\alpha_\ast>0$ such that
for $\alpha$ with $|\alpha|<\alpha_\ast$, $\gsb$ has a ground state $\gr$
such that
$\|(1\otimes N^\han)\gr\|<\infty.$
\ep
\proof See \cite[Theorem 1.3]{ah1}.
\qed
\begin{remark}
In \cite [Theorem 1.3]{ah1}, it is actually supposed that self-adjoint
operator
$A$ has
a compact resolvent, i.e.,
$\sigma(A)=\sigma_{\rm p}(A)$. However it
 can be extended to $A$
satisfying  {\bf (Gap)}.
See \cite[Appendix]{ak}.
\end{remark}
Let $f\in C_0^2(\bk)$ and $\Psi, \Phi \in D(\gsb)$.
We have
\begin{eqnarray*}
[a(f), \gsbbb]_W^{D(\gsb)}(\Psi,\Phi)
&=&
(\Psi,\sum_{j=1}^J(\bar f,\lambda_j)(B_j\otimes 1)\Phi)\\
&=& \int_{\BT} f(k)(\Psi, \sum_{j=1}^J \lambda_j(k) (B_j\otimes 1)\Phi)dk\\
&=&\int_\BT f(k)(\Psi, \tgsb(k) \Phi) dk ,
\end{eqnarray*}
where
$$\tgsb(k):=\sum_{j=1}^J \lambda_j(k) (B_j\otimes 1).$$
Our main theorem in this subsection is as follows.
\bt{sb3}
Suppose {\bf (GSB1)-(GSB3)}, {\bf (IR)},   {\bf (GSB 6)} and {\bf (GSB 7)}.
Then it follows that
\eq{mm1}
P_{\gsb}\FFFg\subset D(1\otimes N^\han),
\en
and
\eq{43}
 \|(1\otimes N^\han)\gr\|^2=
\alpha^2\int_\BT\|(\gsb-E(\gsb)+\ogsb(k))\f\tgsb(k)\gr\|^2 dk.
\en
In addition, suppose {\bf (Gap)}. Then
there exists $\alpha_{\ast\ast}$ such that for $\alpha$ with
$|\alpha|<\alpha_{\ast\ast}$,
it follows that
\eq{mm2}
\ma(\gsb)\leq \ma(A).
\en
\et
\proof
We shall check assumptions {\bf (B1)}-{\bf (B4)} and (3) of Theorem
\ref{asym}
with the following identifications.
$$
\FFF=\FFFg,\ \ \ H_0=\gsbb,\ \ \ \hi=\gsbbb,\ \ \ \omega=\ogsb,\ \ \ D=1,\ \
\ T_{j=1}(k)=\tgsb(k).
$$
{\bf (B1)} and {\bf (B2)} have been  already  checked.
We have
\begin{eqnarray}
&&\int_{\BT}
f(k) (\Psi, \QQ{(\gsb-E(\gsb)+\ogsb (k))}\tgsb(k)\gr)dk \non \\
&&=\int_{\BT}
 f(k) \sum_{j=1}^J\lambda_j(k)\QQ{\ogsb (k)} (\Psi,
\QQ{(\gsb-E(\gsb))} (B_j\otimes 1)
\gr)dk \non \\
&&\label{bound}
=\sum_{j=1}^J
(\Psi, \QQ{(\gsb-E(\gsb))}(B_j\otimes 1) \gr)
\int_{\BT}
 f(k) \lambda_j(k)\QQ{\ogsb (k)}dk.
\end{eqnarray}
Since $f \lambda_j\in C_0^2(\bk)$,
we see that by Lemma \ref{f1},
$$\left|
\int_{\BT} f(k) \lambda_j(k)\QQ{\ogsb (k)} dk\right| \in
L^1([0,\infty),ds),$$
which implies, together with \kak{bound},
that
{\bf (B3)} follows.
We have
$$
\int_\BT \|\tgsb(k)\gr\|^2 dk
\leq
J \sum_{j=1}^J
\lk
\int_\BT |\lambda_j(k)|^2 dk\rk
 \|(B_j\otimes 1)\gr\|^2<\infty,
$$
and
\begin{eqnarray*}
&&\int_{\BT}\|(\gsb-E(\gsb)+\ogsb (k))^{-1} \tgsb(k) \gr\|^2dk\\
&&\leq \int_{\BT}\frac{1}{\ogsb (k)^2}\|\tgsb(k)\gr\|^2dk \\
&&\leq J \sum_{j=1}^J \lk \int_{\BT}\frac{|\lambda_j(k)|^2}{\ogsb (k)^2}
dk\rk
\|(B_j\otimes1)\gr\|^2 <\infty.
\end{eqnarray*}
Thus {\bf (B4)} and (3) of Theorem \ref{asym} follow.
Hence \kak{mm1} and \kak{43} are proven.
We check \kak{mai2} in Theorem~\ref{main2} to show \kak{mm2}.
Note that
\eq{45}
\|(B_j\otimes 1)\gr\|\leq a_j\|(\ov{A}^\han\otimes 1)\gr\|+b_j\|\gr\|
\en
and
$$\|(\ov{A}^\han\otimes 1)\gr\|\leq \|\bhb^\han\gr\|.$$
Since
$$\|\bhb^\han\gr\|^2=(\gr, \bhb\gr)\leq (c_1E(\gsb)+c_2)\|\gr\|^2$$
with some constants $c_1$ and $c_2$, we have
$$\|(B_j\otimes 1)\gr\|\leq (a_j\lk c_1E(\gsb)+c_2\rk^\han  +b_j)\|\gr\|.$$
Thus by \kak{45},
\begin{eqnarray*}
&&
\lim_{g\rightarrow 0}
\sup_{\gr\in P_{\gsb}\FFFg}
\frac{g^2\int_{\BT}\|(\gsb-E(\gsb)+\ogsb (k))^{-1}\tgsb(k)\gr\|^2
dk}{\|\gr\|^2}\\
&& \leq
\lim_{g\rightarrow 0}
g^2 J \sum_{j=1}^J (a_j\lk c_1E(\gsb)+c_2\rk^\han +b_j)^2
\|\lambda_j/\ogsb \|^2
=0.
\end{eqnarray*}
Then \kak{mm2} follows from Theorem \ref{main2}.
\qed

\bc{gsb1}
Assume {\bf (GSB1)}-{\bf (GSB4)}, {\bf (GSB6)}, {\bf (GSB7)}, {\bf (IR)}
and {\bf (Gap)}.
Then there exists
$\alpha_{\ast\ast\ast}$ such that for $\alpha$ with
$|\alpha|<\alpha_{\ast\ast\ast}$,
$\gsb$ has a ground state and $\ma(\gsb)\leq \ma(A)$.
In particular in the case of $\ma(A)=1$, $\gsb$ has a unique ground state.
\ec
\proof
It follows from Proposition \ref{sb2} and Theorem \ref{sb3}.
\qed

\subsection{The Pauli-Fierz model}
The Pauli-Fierz model describes a minimal  interaction
between  electrons with spin $\han$
and a quantized radiation field quantized in the  Coulomb gauge.
The asymptotic field for $\pf$ is studied in e.g.,
\cite{fgs, h12}.
The Hilbert space for state vectors of the Pauli-Fierz Hamiltonian is given
by
$$\FFFp:=L^2(\BD;\CC^2)\otimes \fff(\LW).$$
Formally the annihilation operator and the creation operator of
$\fff(\LW)$
is denoted by
$$\ass(f,j)=\int f(k) \ass(k,j) dk.$$
The Pauli-Fierz Hamiltonian with ultraviolet cutoff $\vp$
is defined by
$$\pf:=\frac{1}{2m}(p\otimes 1-e\av)^2+V\otimes 1+1\otimes \hf-
\frac{e}{2m}(\sigma\otimes1)\cdot  \bv,$$
where
$m>0$ and $e\in \RR$ denote the mass of an electron and the charge of an
electron, respectively.
We regard $e$ as a coupling constant. $p$ denotes the momentum operator of
an electron, i.e.,
$$p=(p_1, p_2, p_3)=(-i \frac{\partial}{\partial x_1},
-i \frac{\partial}{\partial x_2},
-i \frac{\partial}{\partial x_3}),$$
and $V$ is an external potential.
We identify $\FFFp$ as
\eq{id}
\FFFp\cong \CC^2\otimes \int_\BR^\oplus  \fff( \LW) dx,
\en
where $\int^\oplus_\BR\cdots dx$ denotes a constant fiber direct integral
\cite{rs4}.
$\av$ and $\bv$ denote  a quantized radiation field and a quantized magnetic
field
with ultraviolet cutoff $\vp$, respectively, which are given by, under
identification
\kak{id},
$$\av:=1\otimes \int_\BR^\oplus  \av(x) dx, \ \ \
\bv:=1\otimes \int_\BR^\oplus  \bv(x) dx,
$$
with
$$\av(x):=
\jj\int \frac{\vp(k)}{\sqrt{2\opf (k)}}\ekj
\lkk
e^{-ikx} \add(k,j)+
e^{ikx} a(k,j)\rkk dk$$
and
\begin{eqnarray*}
\hspace{-1cm}
\bv(x)&:=&{\rm rot}_x\av(x) \\
\hspace{-1cm}
&=&
\jj\int
\frac{\vp(k)}{\sqrt{2\opf(k)}}(-ik\times  \ekj)
\lkk  e^{-ikx} \add(k,j)-
 e^{ikx} a(k,j)\rkk dk.
\end{eqnarray*}
Here
$$\opf(k):=|k|$$
and $\vp$ denotes an ultraviolet cutoff function.
$$\hf:=\dg{[\opf]}$$ is the second quantization of the multiplication
operator
$$[\opf]:\LW\rightarrow \LW$$
such that 
$$([\opf]f)(k,j)=\opf(k)f(k,j).$$
Vector 
$$\ekj=(e_1(k,j),e_2(k,j), e_3(k,j))\in\BR,\ \ \ j=1,2,$$
denotes a polarization vector satisfying
$$e(k,1)\times e(k,2)=k/|k|,\ \ \ \ |e(k,j)|=1,\ \ \ j=1,2.$$
Note that
$$e(-k,1)=-e(k,1),\ \ \ e(-k,2)=e(k,2).$$
Finally $\sigma:=(\sigma_1,\sigma_2,\sigma_3)$ denotes
  $2\times2$ Pauli matrices
satisfying
the anticommutation relations,
$$\{\sigma_i, \sigma_j\}=2\delta_{ij}, \ \ \
i,j=1,2,3,$$
where $\{A, B\}:=AB+BA$.
Assumptions {\bf (PF1)}- {\bf (PF3)} are  as follows.
\bi
\item
[(PF1)]\
\bi
\item[(1)]
$\sqrt\opf\vp, \vp/\sqrt\opf, \vp/\opf\in\LR$ and
$\vp(k)=\vp(-k)=\ov{\vp(k)}$.
\item[(2)]
$V$ is $\Delta$-bounded with a relative bound strictly less than one.
\eni
\item [(PF2)] \
\bi
\item[(1)] $\vp\in C^\infty(\BD)$.
\item[(2)] $e(\cdot,j)\in C^\infty(\BD\setminus {\cal Q})$, $j=1,2$,
 with some measurable set
${\cal Q}$ with its  Lebesgue measure  zero.
\eni
\item[(PF3)]
The ground state energy of self-adjoint operator
$$\hpp:=-\frac{1}{2m}\Delta+V$$
acting in $\LR$ is discrete.
\eni
Let
$\pff$ be $\pf$ with $e=0$, i.e.,
$$\pff:=\hp \otimes 1+1\otimes \hf,$$
where
$$\hp:=\lk
\begin{array}{cc}\hpp&0\\0&\hpp\end{array}
\rk\mbox{ acting in  } L^2(\BR;\CC^2)\cong \LR\oplus\LR.$$
In what follows, 
simply we write $T\otimes 1$ for $\lk\begin{array}{cc}T&0\\0&T\end{array}\rk\otimes 1$ 
unless confusions arise. 
We note that $\pff$ is self-adjoint on
$$
D(\pff)=D(\Delta \otimes 1)\cap D(1\otimes \hf).
$$
Note that
\eq{N}(p\otimes 1) \cdot \av=\av\cdot (p\otimes 1)
\en 
on $D(\pff)$.
 We set
$$\pf=\pff+e \hip,$$
where, by \kak{N},
$$\hip:=-\frac{1}{m}(p\otimes 1) \cdot\av+
\frac{e}{2m} \av\cdot \av -\frac{1}{2m}(\sigma\otimes1)\cdot   \bv.$$
\bp{hi1}
Assume {\bf (PF1)}, {\bf (PF2)}.
Then $\pf$ is self-adjoint on
$D(\pff)$
and bounded from below. Moreover it is essentially self-adjoint on any core
of $\pff$.
\ep
\proof See \cite{h11,h16}.
\qed
\bp{minna}
Suppose {\bf (PF1)} and {\bf (PF3)}.
Then there exists a constant $e_\ast\leq \infty$ such that
for $e$ with $|e|\leq e_\ast$,
$\pf$ has a ground state such that
$\gr\in D(1\otimes N^\han)$.
\ep
\proof
See e.g., \cite{bfs3, bcv, gll, gl1, gl2,  h8}.
\qed
\begin{remark}
For some $V$, we can take $e_\ast=\infty$. See \cite{gll, gl1}.
\end{remark}
\begin{remark}
Spinless Pauli-Fierz Hamiltonians are defined by
$$\pfs:=\frac{1}{2m}(p\otimes 1-e\av)^2+V\otimes 1+1\otimes \hf,$$
which acts in $\FFF=\LR\otimes \fff(\LW)$.
It can be  proven that
$\pfs$ has a ground state $\gr$ such that $\gr\in D(1\otimes N^\han)$,
and it is unique \cite{h9}.
Then it follows that
$P_\pfs \FFF\subset D(1\otimes N^\han)$. 
\end{remark}
We have
\begin{eqnarray*}
&& [1\otimes a(f,j), \hip]_W^{D(\pf)} (\Psi,\Phi)\\
&&=
(\Psi, \lkk -\frac{1}{m}\kk(x,j)\cdot (p\otimes 1-e\av)-\frac{1}{2m}
\wk(x,j)\cdot (\sigma\otimes 1) \rkk \Phi),
\end{eqnarray*}
where
\begin{eqnarray*}
\kk(x,j):=(\ov f, \lambda_j^{(1)}(x) ),\ \ \
\wk(x,j):=(\ov f, \lambda_j^{(2)}(x) ),
\end{eqnarray*}
and
\begin{eqnarray*}
\lambda_j^{(1)}(x,k)&:=&\mmm,\\
 \lambda_j^{(2)}(x,k)&:=&\mmmm.
\end{eqnarray*}
Hence we have
$$
[1\otimes a(f,j), \hip]_W^{D(\pf)} (\Psi,\Phi)=\int f(k) (\Psi,
\tpf_j(k)\Psi) dk,$$
where
\begin{eqnarray*}
&& \tpf_j(k):=\tpf_j^{(1)}(k)+\tpf_j^{(2)}(k),\\
&& \tpf_j^{(1)}(k):= -\frac{1}{m}\mmm \cdot (p\otimes 1-e\av),\\
&& \tpf_j^{(2)}(k):= -\frac{1}{2m} \mmmm \cdot (\sigma\otimes 1).
\end{eqnarray*}
Let $\pfz$ be $\pf$ with $V=0$.
Then the binding energy is defined by
$$\eb:=E(\pfz)-E(\pf).$$
\bp{decay}
Assume that $\hpp$
has a ground state in $\LR$.
Then $$\eb\geq -E(\hpp).$$
\ep
\proof
See Appendix \ref{53}.
\qed
Assumption {\bf (V)} is as follows.
\bi
\item[(V)]
Potential $V=V_+-V_-$ ($V_+(x)=\max\{0, V(x)\}$, $V_-(x)=\min\{0,V(x)\}$)
satisfies that
(1) $\lim_{|x|\rightarrow \infty} V_-(x)=V_\infty<\infty$,
(2) $|x|^2 V_-\in L_{\rm loc}^\infty(\BR)$,
(3) $\eb>V_\infty$.
\eni
A typical example of $V$ in {\bf (V)} is Coulomb potential
$-e^2/|x|$.
\bl{lo}
Suppose  {\bf (V)}. Then it follows that
$$\sup_{\gr\in \pg\FFF}\frac{\|(|x|\otimes 1)\gr\|}{\|\gr\|}<\cex$$
with some constant $\cex$.
\el
\proof
See Appendix \ref{54}.
\qed
The main theorem in this subsection is as follows.
\bt{mainhiro}
Assume {\bf (PF1)},  {\bf (PF2)},  and  {\bf (V)}.
Then it follows that
\eq{pp1}
P_\pf\FFFp\subset D(1\otimes N^\han),
\en
and
\eq{pp3}
\|(1\otimes
N^\han)\gr\|^2=
e^2\jj\int_\BR\|(\pf-E(\pf)+\opf(k))^{-1}\tpf_j(k)\gr\|^2 dk.
\en
In addition, assume {\bf (PF3)}, then
there exists a constant $e_{\ast\ast}$
such that for $e$ with $|e|<e_{\ast\ast}$,
\eq{pp2}
\ma(\pf)\leq \ma(H_{\rm p}). 
\en
\et
\begin{remark}
Suppose that e.g.,
$V_+\in L_{\rm loc}^1(\BR)$ and
$V_-$ is infinitesimally small with respect to $\Delta$.
Then, by a Feynman-Kac formula,
it is shown that $e^{-t(\hpp-E(\hpp))}$ is positivity improving in $\LR$.
Hence $\hpp$
has a unique ground state in $\LR$.
Then
$\ma(H_{\rm p})=2$.
\end{remark}
To prove Theorem \ref{mainhiro}
it is sufficient to check ${\bf (B1)}$-${\bf (B4)}$ and (3) of Theorem
\ref{asym}
with the following identifications.
$$\FFF=\FFFp,\ \ \ H_0=\pff,\ \ \ \hi=\hip,\ \ \ \omega=\opf,\ \ \ D=2,\ \ \
T_j(k)=\tpf_j(k).$$
Let
$$K:=\bigcup_{n=1}^3 \{(k_1,k_2,k_3)\in\BR| k_n=0\}$$
and
$$\widetilde{K}:=K\cup {\cal Q} \cup\{0\}.$$
\bl{hi3}
Assume {\bf (PF1)} and  {\bf (PF2)}. Then
for $f\in C_0^2(\BD\setminus \widetilde{K})$ and $\Psi\in D(\pf)$,
$$\left|
\int_{\BT}
f(k)(\Psi, \QQ{(\pf-E(\pf)+\opf(k))} \tpf_j^{(l)}(k) \gr) dk \right|\in
L^1([0,\infty),ds), \ \ \ l=1,2.$$
\el
\proof
Note that
$$\ooo [\kk_\mu(x,j),   (p\otimes 1-e\av)_\mu]=0$$
on $D(\pf)$.
We see that
\begin{eqnarray*}
&&\int_{\BT}
f(k)(\Psi, \QQ{(\pf-E(\pf)+\opf(k))} \tpf_j^{(1)}(k) \gr)dk \\
&&=-\frac{1}{m}\ooo ((p\otimes 1-e\av)_\mu \QQ{(\pf-E(\pf))}\Psi,
\kk_\mu (s,x,j)\gr),
\end{eqnarray*}
where
$$\kk(s,x,j):=(\Q\opf \bar f, \lambda_j^{(1)}(x)).$$
Since
$$\Q\opf=
\frac{\opf(k)}{k_\mu}\frac{1}{is}\frac{\partial}{\partial k_\mu}\Q \opf,\ \
\ \mu=1,2,3,$$
we have
$$\kk(s,x,j)=\frac{1}{s}(\kk_1(s,x,j)+x_\mu \kk_2(s,x,j)),$$
where
\begin{eqnarray*}
&&\kk_1(s,x,j):=
-i\int_\BD e^{-i(s\opf(k)+kx)}
\frac{\partial}{\partial k_\mu}
\lk \frac{\opf(k)}{k_\mu}
\frac{\vp(k)}{\sqrt{2\opf(k)}}
f(k)\ekj \rk dk, \\
&&\kk_2(s,x,j):=
\int_\BD e^{-i(s\opf(k)+kx)} \frac{\partial}{\partial k_\mu}\lk
\frac{\vp(k)}{\sqrt{2\opf(k)}}f(k)\ekj\rk dk.
\end{eqnarray*}
From the fact that
$\vp\in C^\infty(\BD)$ and $f\in C_0^2(\BD\setminus \wwk)$,
it follows that for $\nu=1,2,3$,
\begin{eqnarray*}
&& \frac{\partial}{\partial k_\mu}\lk
\frac{\opf(k)}{k_\mu}\frac{\vp(k)}{\sqrt{2\opf(k)}}
f(k)e_\nu(k,j)\rk \in C_0^\infty(\BD\setminus\{0\}),\\
&&
\frac{\partial}{\partial k_\mu}\lk
\frac{\vp(k)}{\sqrt{2\opf(k)}}
f(k)e_\nu(k,j) \rk \in C_0^\infty(\BD\setminus\{0\}).
\end{eqnarray*}
Thus by \cite[Theorem XI.19 (c)] {rs3}
there exist constants $c_1$ and $c_2$ such that
\eq{hi4}
\sup_x|\kk_{l,\mu}(s,x,j)|\leq \frac{c_l}{1+s},\ \ \ l=1,2, \ \ \ \lll.
\en
By this we have
\eq{hi5}
\|K^{(1)}_\mu(s,x,j)\gr\|\leq
\frac{1}{s(s+1)}(c_1\|\gr\|+c_2\|(|x|\otimes 1) \gr\|).
\en
Since
\eq{jisin}
\|(p\otimes 1-e\av)_\mu\Psi\|\leq c_1'\|(\pf-E(\pf))\Psi\|+c_2'\|\Psi\|
\en
with some constants $c_1'$ and $c_2'$,
we conclude that
\begin{eqnarray}
&&
\hspace{-1cm}
\left|
-\frac{1}{m}\ooo
((p\otimes 1-e\av)_\mu
\QQ{(\pf-E(\pf))}\Psi,
K_\mu(s,x,j) \gr)
\right|\non\\
\label{hi7}
&&
\hspace{-1cm}
\leq
\frac{3}{m}(c_1'\|(\pf-E(\pf))\Psi\|+c_2'\|\Psi\|)
(c_1\|\gr\|+c_2\|(|x|\otimes 1)  \gr\|)\frac{1}{s(1+s)}.\ \ \ \ \ \ \ \ \ \
\end{eqnarray}
From this it follows that
$$-\frac{1}{m}\ooo ((p\otimes 1-e\av)_\mu
\QQ{(\pf-E(\pf))}\Psi, K_\mu(s,x,j)\gr)\in
L^1([0,\infty),ds).$$
Similarly we can estimate
\begin{eqnarray*}
&&
\int_{\BT}
f(k)(\Psi, \QQ{(\pf-E(\pf)+\opf(k))} \tpf_j^{(2)}(k) \gr) dk \\
&&=-\frac{1}{2m}\ooo ((\sigma_\mu \otimes 1)  \QQ{(\pf-E(\pf))}\Psi,
\wk_\mu(s,x,j) \gr),
\end{eqnarray*}
where
$$\wk(s,x,j):=(\Q\opf \bar f, \lambda^{(2)}_j(x)).$$
We have
$$\wk(s,x,j)=\frac{1}{s}(\wk_1(s,x,j)+x_\mu \wk_2(s,x,j)),$$
where
\begin{eqnarray*}
&&\wk_1(s,x,j)\\
&&:=
-i\int_\BD e^{-i(s\opf(k)+kx)}
\frac{\partial}{\partial k_\mu}\lk \frac{\opf(k)}{k_\mu}
\frac{\vp(k)}{\sqrt{2\opf(k)}}f(k)(-ik\times \ekj)\rk dk, \\
&&\wk_2(s,x,j)\\
&&:=
\int_\BD e^{-i(s\opf(k)+kx)}
 \frac{\partial}{\partial k_\mu}\lk
\frac{\vp(k)}{\sqrt{2\opf(k)}}f(k)(-ik\times \ekj)\rk dk.
\end{eqnarray*}
Since for $\mu, \nu=1,2,3$,
\begin{eqnarray*}
&&\frac{\partial}{\partial k_\mu}\lk \frac{\opf(k)}{k_\mu}
\frac{\vp(k)}{\sqrt{2\opf(k)}}f(k)(-ik\times \ekj)_\nu\rk\in
C_0^\infty(\BT\setminus\{0\}), \\
&&
\frac{\partial}{\partial k_\mu}\lk
\frac{\vp(k)}{\sqrt{2\opf(k)}}f(k)(-ik\times \ekj)_\nu\rk\in
C_0^\infty(\BT\setminus\{0\}),
\end{eqnarray*}
we can see
that there exist constants $\widetilde{c}_1$ and $\widetilde{c}_2$ such that
\eq{hi6}
\sup_x|K_{l,\mu}^{(2)}(s,x,j)|\leq\frac{\widetilde{c}_l}{1+s}, \ \ \ l=1,2,\
\ \ \lll.
\en
Then we conclude that
\begin{eqnarray*}
&&\left|
-\frac{1}{2m}\ooo ((\sigma_\mu \otimes1) \QQ{(\pf-E(\pf))}\Psi,
\wk_\mu(s,x,j)\gr)\right|\\
&&
\leq
\frac{3}{2m} \| \Psi\|\lk
\widetilde{c}_1
\|\gr\|+\widetilde{c}_2\|(|x|\otimes 1)  \gr\|\rk\frac{1}{s(1+s)}.
\end{eqnarray*}
Thus
$$-\frac{1}{2m}\ooo ((\sigma_\mu \otimes1) \QQ{(\pf-E(\pf))}\Psi,
\wk_\mu(s,x,j) \gr)\in L^1([0,\infty),ds).$$
Hence the lemma is proven.
\qed
\bl{hi8}
We have
$$\|\tpf^{(l)}_j(\cdot)\gr\|\in \LR,\ \ \ l=1,2.$$
\el
\proof
It follows that
\begin{eqnarray*}
\|\tpf^{(1)}_j(k)\gr\|
&\leq&
\frac{1}{m}\ooo
\frac{|\vp(k)| }{\sqrt{2\opf(k)}}
|\ekj_\mu |
\|(p\otimes 1-e\av)_\mu\gr\|\\
&
\leq &
\sum_{\mu=1}^3
\frac{1}{m}
\frac{|\vp(k)|}{\sqrt{2\opf(k)}}
\|(p\otimes 1-e\av)_\mu\gr\|,
\end{eqnarray*}
and
\begin{eqnarray*}
\|\tpf^{(2)}_j(k)\gr\|
&\leq&
\frac{1}{2m}\ooo
\frac{|\vp(k)|}{\sqrt{2\opf(k)}}
|(k\times \ekj)_\mu| \|(\sigma_\mu\otimes1)\gr\|\\
&
\leq &
\frac{3}{2m} \frac{\vp(k)}{\sqrt{2\opf(k)}}
|k| \|\gr\|.
\end{eqnarray*}
Since $\sqrt\opf\vp,\vp/\sqrt\opf\in\LR$,  the lemma follows.
\qed
\bl{hi0}
Suppose that $\Psi\in D(\pff)\cap D(|x|\otimes 1)$ and
$\pf \Psi\in D(|x|\otimes 1)$. Then for $\mu=1,2,3$,
\bi
\item[(1)]
$(x_\mu\otimes1) \Psi\in D(\pf)$,
\item[(2)]
$\displaystyle [x_\mu\otimes 1, \pf] \Psi=\frac{i}{m}(p\otimes 1-e\av)_\mu
\Psi.$
\eni
In particular, it follows that $(x_\mu\otimes 1)\gr\in \pf$ with
$$\frac{i}{m}(p\otimes 1-e\av)_\mu \gr
=[x_\mu\otimes 1,  \pf]\gr
= (\pf-E(\pf)) (x_\mu\otimes 1) \gr.$$
\el
\proof
See Appendix \ref{gu}.
\qed
\bl{hi9}
Assume {\bf (PF1)}. Then
$$\int_\BD \|(\pf-E(\pf)+\opf(k))^{-1}\tpf_j^{(l)}(k)\gr\|^2dk<\infty,\ \ \
l=1,2.$$
\el
\proof
Note that by Lemma \ref{hi0}
\begin{eqnarray*}
&& \tpf_j^{(1)}(k)\gr\\
&&= -\frac{1}{m}\mmm \cdot (p\otimes 1-e\av)\gr\\
&&=-\frac{1}{m}\mmm \cdot (-im )[x\otimes 1,\pf]\gr\\
&&=
-\frac{1}{m}\mmm \cdot  (-im) (\pf-E(\pf)) (x\otimes 1)\gr.
\end{eqnarray*}
Hence we have
\begin{eqnarray*}
&&\hspace{-0.5cm}
 (\pf-E(\pf)+\opf(k))^{-1}\tpf_j^{(1)}(k)\gr\\
&&\hspace{-0.5cm}
=i \frac{\vp(k)}{\sqrt{2\opf(k)}} \ekj
(\pf-E(\pf)+\opf(k))^{-1}e^{-ikx}(\pf-E(\pf))
(x\otimes1)\gr\\
&&\hspace{-0.5cm}
=i \frac{\vp(k)}{\sqrt{2\opf(k)}} \ekj (\pf-E(\pf)+\opf(k))^{-1}\\
&& \hspace{3cm} \times (\pf(k)-E(\pf))e^{-ikx}
(x\otimes 1)\gr,
\end{eqnarray*}
where we used that
$e^{ikx}$ maps $D(\pff)$ onto itself (see  Appendix \ref{gu}) and on
$D(\pf)$,
\begin{eqnarray*}
&& \pf(k):=e^{-ikx}\pf e^{ikx}\\
&&=
\frac{1}{2m}(p\otimes 1+k-e\av)^2+V\otimes 1+1\otimes
\hf-\frac{1}{2m}(\sigma\otimes 1)
\cdot\bv\\
&&= \pf +\frac{1}{m}(p\otimes 1-e\av)\cdot k+\frac{1}{2m}|k|^2.
\end{eqnarray*}
Thus we have
\begin{eqnarray}
&& \| (\pf-E(\pf)+\opf(k))^{-1}\lk \pf(k)-E(\pf)\rk
e^{-ikx}(x_\mu\otimes1)\gr\|\non \\
&&
\label{hi11}
\leq
\| (\pf-E(\pf)+\opf(k))^{-1}(\pf-E(\pf))
e^{-ikx}(x_\mu\otimes 1)\gr\|\\
&&
\label{hi12}
+\| (\pf-E(\pf)+\opf(k))^{-1}
\frac{1}{m}(p\otimes 1-e\av)\cdot k  e^{-ikx}(x_\mu\otimes 1)\gr\|\\
&&
\label{hi13}+\| (\pf-E(\pf)+\opf(k))^{-1}
 \frac{1}{2m}|k|^2  e^{-ikx}(x_\mu\otimes 1)\gr\|.
\end{eqnarray}
We have
\eq{hi14}
|\kak{hi11}|\leq \|(|x|\otimes 1)\gr\|
\en
and
\eq{hi15}
|\kak{hi13}|\leq \frac{1}{2m}\frac{|k|^2}{\opf(k)} \|(|x|\otimes 1)\gr\|=
\frac{1}{2m}|k|\|(|x|\otimes1)\gr\|.
\en
Note that by \kak{jisin},
\begin{eqnarray*}
&& \|(p\otimes 1-e\av)_\mu(\pf-E(\pf)+\opf(k))^{-1}\gr\|\\
&&\leq
c_1' \|(\pf-E(\pf))(\pf-E(\pf)+\opf(k))^{-1}\gr\|\\
&&\hspace{1cm}
+c_2'\|(\pf-E(\pf)+\opf(k))^{-1}\gr\|\\
&&\leq c_1' \|\gr\|+\frac{c_2'}{\opf(k)} \|\gr\|.
\end{eqnarray*}
Then
\eq{hi16}
|\kak{hi12} | \leq \frac{3}{m}|k|(c_1'+\frac{c_2'}{\opf(k)})
\|(|x|\otimes 1)\gr\|=\frac{3}{m}(c_1'|k|+c_2')\|(|x|\otimes 1) \gr\|.
\en
Together with \kak{hi14}, \kak{hi15}, \kak{hi16},  we have
\begin{eqnarray}
&& \|(\pf-E(\pf)+\opf(k))^{-1}\tpf_j^{(1)}(k)\gr\|\non \\
&&\label{qq11}
\leq
3 \frac{|\vp(k)|}{\sqrt{2\opf(k)}}\lk1+\frac{|k|}{2m}+
\frac{3}{m}(c_1'|k|+c_2')
\rk \|(|x|\otimes 1)\gr\|.
\end{eqnarray}
Since
$\sqrt\opf\vp, \vp/\sqrt\opf\in\LR$,
\eq{qq1}
\int_\BD \|(\pf-E(\pf)+\opf(k))^{-1}\tpf_j^{(1)}(k)\gr\|^2dk<\infty
\en
follows.
Moreover we have
\begin{eqnarray}
&&\|(\pf-E(\pf)+\opf(k))^{-1}\tpf_j^{(2)}(k)\gr\|\non\\
&&\leq
\frac{3}{2m}
\frac{|\vp(k)|}{\sqrt{2\opf(k)}}|k|\frac{1}{\opf(k)}\|\gr\|\non \\
\label{qq22}
&&=\frac{3}{2m}\frac{|\vp(k)|}{\sqrt{2\opf(k)}}\|\gr\|.
\end{eqnarray}
Hence
\eq{qq2}
\int_\BD \|(\pf-E(\pf)+\opf(k))^{-1}\tpf_j^{(2)}(k)\gr\|^2dk<\infty
\en
follows.
Thus by \kak{qq1} and \kak{qq2},
we get the desired results.
\qed

{\it Proof of Theorem \ref{mainhiro}}\\
Lemmas \ref{hi3}, \ref{hi8} and  \ref{hi9} correspond  to
assumptions {\bf (B3)}, {\bf (B4)} and (3) of Theorem \ref{asym},
respectively.
Then \kak{pp1} and \kak{pp3} follow from Theorem \ref{asym}.
By \kak{qq11} and \kak{qq22},  we have
\begin{eqnarray*}
&&\frac{\|(1\otimes N^\han)\gr\|^2}{\|\gr\|^2}\\
&& =
e^2 \frac{\sum_{j=1,2}
\int_\BR\|(\pf-E(\pf)+\opf(k))^{-1}\tpf(k)^{(j)}\gr\|^2 dk}{\|\gr\|^2}\\
&&\leq
6e^2 \int\lkk
3\frac{|\vp(k)|}{\sqrt{2\opf(k)}}\lk1+\frac{|k|}{2m}+
\frac{3}{m}(c_1'|k|+c_2')\rk\rkk^2 dk
\frac{\|(|x|\otimes 1)\gr\|^2}{\|\gr\|^2}\\
&&\hspace{2cm}+ 6e^2 \int \lkk
\frac{3}{2m}\frac{|\vp(k)|}{\sqrt{2\opf(k)}}\rkk^2 dk.
\end{eqnarray*}
Since, by Lemma \ref{exp},
$$\frac{\|(|x|\otimes 1)\gr\|^2}{\|\gr\|^2}\leq \cex,$$
we obtain
$$
\lim_{e\rightarrow0}
\sup_{\gr\in P_{\pf}\FFFp}
e^2 \frac{\sum_{j=1,2}
\int_\BR\|(\pf-E(\pf)+\opf(k))^{-1}\tpf(k)^{(j)} \gr\|^2 dk}{\|\gr\|^2}=0.
$$
Thus
\kak{pp2} follows from Theorem \ref{main2}.
\qed
\begin{remark}
Although,  in \cite{hisp2},
formula \kak{pp3}:
$$\|(1\otimes N^\han)\gr\|^2=
e^2 \jj \int_\BR \|(\pf-E(\pf)+\opf(k))\f
\tpf_j(k)\|^2 dk$$
has been used to show $\ma(\hp)\leq 2$,
there is no exact proof to derive this formula in it.
See Subsection \ref{plo}.
\end{remark}
In \cite{hisp2} it has been  also proven that $2\leq \ma(\pf)$ under some
conditions on $V$. 
We state a theorem. 
\bt{hisp}
In addition to  ${\bf (PF1)}-{\bf (PF3)}$  and ${\bf (V)}$, 
we assume $\ma(\hp)=2$ and $V(x)=V(-x)$. Then 
there exists a constant $e_{\ast\ast\ast}$ such that 
for $e$ with $|e|<e_{\ast\ast\ast}$, $\ma(\pf)=2$.
\et
\proof 
$\ma(\pf)\leq 2$ follows from Theorem \ref{mainhiro} 
and $2\leq \ma(\pf)$ from \cite{hisp2}.
We omit details. 
\qed

\subsection{Concluding remarks}
We can apply the method stated in this paper
to a wide class of interaction Hamiltonians in quantum field models.
 Hamiltonian $\cd$  of the Coulomb-Dirac system  is defined as
an operator
acting in
$$\FFF_{\rm CD}={\cal F}_{\rm f}(\oplus^4 \LR)\otimes
\fff(\LW),$$
where ${\cal F}_{\rm f}(\oplus^4\LR)$ denotes a fermion Fock space over
$\oplus^4\LR$.
The Coulomb-Dirac system describes an interaction of positrons and
relativistic electrons
through photons in the Coulomb gauge.
Operator $\cd$  is of the form
$$\cd=H_{\rm f}\otimes 1+1\otimes \dg{\omega_{\rm CD}}+e\hicd+e^2\hiicd,$$
where $H_{\rm f}$ denotes a free Hamiltonian of ${\cal F}_{\rm
f}(\oplus^4\LR)$,
$\omega_{\rm CD}$ the multiplication operator by $\omega_{\rm CD}(k)= |k|$,
and
$\hicd$, $\hiicd$  interaction terms.
$\cd$ has been investigated in
\cite{  bdg3}, where the self-adjointness and the existence of a
ground state
are proven under some conditions.
It is known that $\ma(H_{\rm f})=1$.
Then
using the method in this paper we can also show
$$\ma(\cd)\leq\ma(H_{\rm f})=1,$$
i.e.,
the ground state of $\cd$ is unique for a sufficiently small $e$.
We omit details.

\section{Appendix}
\subsection{Proofs of \kak{29} and  \kak{31}}
\begin{proposition}[\kak{29} and \kak{31}]\label{p}
We have
$\gr\in D(\bh^\han)$, $Q\gr\in D(\bh^\han)$ and
\eq{41}
\bh^\han Q\gr=Q\bh^\han \gr.
\en
\end{proposition}
\proof
From the fact that the form domain $Q(H)$ of $H$ satisfies
$$ D(H)\subset Q(H)=D(\bh^\han),$$
it follows that $\gr\in D(\bh^\han)$.
Since
$[e^{-t\bh^\han}, Q]=0$,
we have
\eq{gan}
\frac{e^{-t\bh^\han}-1}{t} Q\gr=Q\frac{e^{-t\bh^\han}-1}{t} \gr,\ \ \ t>0.
\en
Since $\gr\in D(\bh^\han)$,
we see that
$$s-\lim_{t\rightarrow 0}
Q\frac{e^{-t\bh^\han}-1}{t} \gr=
-Q\bh^\han\gr.$$
Hence  the left-hand side of  \kak{gan} converges, which
implies  that
$Q\gr\in D(\bh^\han)$ and \kak{41} follows.
\qed

\subsection{Proof of \kak{chi}}
\begin{proposition}[\kak{chi}]\label{ko-taro-}
Assume that
\eq{ass}
\lim_{g\rightarrow 0}\sup_{\Psi\in
D(\bh^\han)}\frac{|\beta(\Psi,\Psi)-\beta_0(\Psi,\Psi)|}{\|(\bh+1)^\han\Psi\
|^2}=0.
\en
Then for $z\in\CC$ with $\Im z\not=0$,
$$\lim_{g\rightarrow 0}
\|(\hq-z)\f-(\bh-z)\f\|=0.$$
\end{proposition}
\proof
Set
$$\J:=(\bh+1)^{-\han}.$$
We have
\begin{eqnarray}
&& (\Psi, \J \hq  \J \Psi)\nonumber \\
&&=(\Psi, \J(\hq _+^\han \hq _+^\han -\hq _-^\han \hq
_-^\han)\J\Psi)\nonumber \\
&&= (\Psi, \J(\hq _+^\han \hq _+^\han)\J \Psi)-(\Psi, \J(\hq _-^\han
\hq _-^\han)\J\Psi)\nonumber \\
&&=(\hq _+^\han\J\Psi,\hq _+^\han\J\Psi)-(\hq _-^\han\J\Psi,\hq
_-^\han\J\Psi)\nonumber
\\
&&\label{215}=\beta(\J\Psi,\J\Psi) \\
&&\leq (1+|g|a) \beta_0(\J \Psi, \J \Psi)+|g| b(\J \Psi,\J \Psi)\nonumber \\
&& \leq C(g) \|\Psi\|^2,\nonumber
\end{eqnarray}
where
$$C(g):=1+|g|(a+b).$$
Then
$\J \hq \J$ is a bounded operator and
 $$\|\J \hq \J\|<C(g).$$
Since the range of $\J$ equals to $D(\bh^\han)$, we have from \kak{ass}
\eq{216}
\lim_{g\rightarrow 0}\sup_{\Psi\in
\FFF}\frac{|\beta(\J\Psi,\J\Psi)-\beta_0(\J\Psi,
\J\Psi)|}{\|\Psi\|^2}=0.
\en
By \kak{215}, i.e.,
$$\beta(\J\Psi,\J\Psi)=(\Psi, \J \hq \J\Psi)$$ and
$$\beta_0(\J\Psi,\J\Psi)=(\Psi, \J \bh\J\Psi),$$
\kak{216} implies that
$$
\lim_{g\rightarrow 0}\sup_{\Psi\in \FFF}\frac{|(\Psi, \J(\hq -\bh )\J\Psi)|}
{\|\Psi\|^2}=0.$$
Hence
we obtain that
\eq{217}
\lim_{g\rightarrow 0}\|\J(\hq -\bh )\J\|=0.
\en
Moreover, for $z\in\CC$ with $\Im z\not=0$,
\begin{eqnarray*}
&&|\beta_0(\JJ\Psi,\JJ\Psi)|\\
&&\leq |\beta(\JJ\Psi,\JJ\Psi)|+|g\betai(\JJ\Psi,\JJ\Psi)|\\
&&\leq |(\JJ\Psi, \hq \JJ\Psi)|+a|g|\beta_0(\JJ\Psi,\JJ\Psi)\\
&&\hspace{3cm} +b|g|(\JJ\Psi,\JJ\Psi).
\end{eqnarray*}
Then
\begin{eqnarray*}
&&\|\bh^\han \JJ\Psi\|^2\\
&&= \beta_0(\JJ\Psi,\JJ\Psi)\\
&&\leq\frac{1}{1-a|g|}
\lkk\lk \JJ\Psi, \hq \JJ\Psi\rk\right.\\
&&\left. \hspace{4cm}+b|g|
\lk \JJ\Psi,\JJ\Psi\rk\rkk\\
&&=\int_{\is(\hq )}^\infty
\lk
\frac{1}{1-|g|a}\frac{\lambda+|g|b}{(\lambda-z)^2}\rk
d\|E_\hq (\lambda)\Psi\|^2\\
&&\leq d(g)\|\Psi\|^2 ,
\end{eqnarray*}
where
$$d(g):=\sup_{\lambda\in\RR}
\left|
\frac{1}{1-|g|a}\frac{\lambda+|g|b}{(\lambda-z)^2}\right|.$$
Thus we see that
$$\|K_0\f  \JJ\|<\|\bh ^\han\JJ\|+\|\JJ\|\leq \sqrt{d(g)}+\frac{1}{|\Im
z|}:=D(g).$$
Directly we have
\begin{eqnarray*}
&&\|(\hq -z)\f-(\bh -z)\f\|\\
&&\leq
\|(\hq -z)\f K_0\f \|
\|\J (\hq -\bh )\J\|
\|K_0\f (\bh -z)\f  \| \\
&&\leq
 D(g)D(0)  \|\J (\hq -\bh )\J\|.
\end{eqnarray*}
By \kak{217} and the fact
$$\lim_{g\rightarrow 0} D(g)<\infty,$$
we obtain that
$$\lim_{g\rightarrow 0}\|(\hq -z)\f-(\bh -z)\f\|=\lim_{g\rightarrow 0}
D(g)D(0)
\|\J (\hq -\bh )\J\|=0.$$
Thus the proposition is proven.
\qed

\subsection{Proof of Proposition \ref{decay}}
\label{53}
{\it Proof of Proposition \ref{decay}}

We see an outline of a proof. See \cite{gll,hhs} for  details.
It is enough to show that for an arbitrary $\epsilon>0$,
 there exists a vector $\Psi_\epsilon \in D(\pf)$ such that
\eq{lpp}
(\Psi_\epsilon, (\pf-(E(\pfz)+\epsilon+E(\hpp))\Psi_\epsilon)<0.
\en
We identify  $\FFFp$ with $L^2(\BR; \fffc)$, i.e., $\fffc$-valued
$L^2$-function over $\BR$.
Let $f$ be a ground state of  $\hpp$. Assume that $f$ is real
and $\|f\|=1$.
For $\epsilon>0$, let
\eq{er}
\Phi_\epsilon=\sum^{\rm finite}_m f_m^\epsilon
\otimes \Phi_m^\epsilon
\en
be such that
$f_m^\epsilon \in C_0^\infty(\BR)$, $\Phi_m^\epsilon \in\oplus^2\ffff$ and
\eq{er2}
(\Phi_\epsilon, \pfz\Phi_\epsilon)\leq E(\pfz)+\epsilon.
\en
Actually, since the linear hull of vectors such as \kak{er} is a core of
$\pf$,
there exists $\Phi_\epsilon$ such as \kak{er} satisfying \kak{er2}.
Note that
$f\Phi_\epsilon\in\FFF$.
Let
$$U_y:=e^{iy\cdot p}\otimes 1,\ \ \ y\in\BR.$$
We can see that $fU_y\Phi_\epsilon \in D(\pf)$ and
\begin{eqnarray*}
\Omega_y&:=&
(fU_y\Phi_\epsilon,
(\pf-(E(\pfz)+\epsilon+E(\hpp))fU_y\Phi_\epsilon)_{\FFFp}\\
&=&
\int_\BR\lkk (\Phi_\epsilon(x),
(\pfz\Phi_\epsilon)(x))_{\fffc}-(E(\pfz)+\epsilon)\|\Phi_\epsilon(x)
\|_\fffc^2\rkk f(x-y)^2 dx\\
&&
+\int_\BR(\Phi_\epsilon(x), ((p\otimes
1)\Phi_\epsilon)(x))_\fffc f(x-y)(pf)(x-y)dx.
\end{eqnarray*}
Note that $fU_y\Phi_\epsilon$ and $\pff fU_y\Phi_\epsilon$ are strongly
continuous in $y$.
Then $\Omega_y$ is continuous in $y$.
We have
\begin{eqnarray*}
&&\int_\BR\Omega_y dy\\
&&\hspace{-0.2cm}=
\int_\BR dy
\int_\BR dx \lkk (\Phi_\epsilon(x),
(\pfz\Phi_\epsilon)(x))_{\fffc}-(E(\pfz)+\epsilon)
\|\Phi_\epsilon(x)\|_\fffc^2\rkk f(x-y)^2 \\
&&\hspace{-0.2cm}=
\int_\BR dx \lkk (\Phi_\epsilon(x),
(\pfz\Phi_\epsilon)(x))_{\fffc}-(E(\pfz)+\epsilon)
\|\Phi_\epsilon(x)\|_\fffc^2\rkk \\
&&\hspace{-0.2cm}=(\Phi_\epsilon, (\pfz-(E(\pfz)+\epsilon)\Phi_\epsilon)<0.
\end{eqnarray*}
Here we used that $\|f\|=1$,  and by the fact that $f$ is real,
$$\int_\BR f(x-y)(p_\mu f)(x-y)dy=(f, p_\mu f )=0,\ \ \ \mu=1,2,3.$$
Thus we conclude that
there exists $y_0\in\BR$ such that
$\Omega_{y_0}<0$, which yields that
$$(fU_{y_0}\Phi_\epsilon,
(\pf-(E(\pfz)+\epsilon+E(\hpp))fU_{y_0}\Phi_\epsilon)<0.$$
The proof is complete.
\qed

\subsection{Proof of Lemma \ref{lo}}
\label{54}
We show a more general lemma than Lemma \ref{lo}.
\bl{exp}
Assume the following facts.
\bi
\item[(1)] $\lim_{|x|\rightarrow \infty}V_-(x)=V_\infty<\infty$.
\item[(2)] A positive function $G$ satisfies that
\bi
\item[(i)] $G\in C^1(\BR\setminus {\cal N})\cap C(\BR)$ with some compact
set ${\cal N}$,
\item[(ii)] $\sup_{x\in\BR\setminus {\cal N}} |\nabla G(x)|<\infty$,
\item[(iii)] $ G^2V_- \in L_{\rm loc}^\infty(\BR)$,
\item [(iv)] $\lim_{|x|\rightarrow \infty}|G(x)||x|^{-l}=d<\infty$ with some
$l\geq 0$ and $d>0$.
\eni
\item[(3)] $\eb>V_\infty$.
\eni
Then there exists a constant $\cex$ such that
\eq{de}
\sup_{\gr\in \pgf\FFF}\frac{\|(G\otimes 1)\gr\|}{\|\gr\|}\leq \cex.
\en
\el
\proof
This proof is a generalization of \cite{gll}.
Let $\chi_{|x|>R}\in C^\infty(\BR)$ be a function such that
$\chi_{|x|>R}(x)=0$ for
 $|x|<R$, and $\chi_{|x|>R}(x)=1$ for $|x|>R+1$.
Since ${\cal N}$ is compact,
$\chi_{|x|>R} G \in C^1(\BR)$
for
a sufficiently large $R$.
Since $(1- \chi_{|x|>R})G$ is a bounded operator,
it is enough to prove \kak{de} with $G$ replaced by $\chi_{|x|>R}G$.
We reset $\chi_{|x|>R}G$ as $G$.

(1)
Suppose that $G\in C_0^\infty(\BR)$. Then we have
$$(G\gr, (\pf-E(\pf))G \gr)=\frac{1}{2m}(\gr, |\nabla G|^2\gr).$$
The left-hand side above is
\begin{eqnarray*}
(G\gr, (\pf-E(\pf))G \gr)
&=& (G\gr, \lkk (\pfz-E(\pf))+V \rkk G\gr)\\
&\geq&
 \eb \|G\gr\|^2+(G\gr, VG\gr)\\
&\geq &
 \eb \|G\gr\|^2+(G\gr, -V_-G\gr)..
\end{eqnarray*}
Hence
\eq{441}
\eb \|G\gr\|^2\leq \frac{1}{2m}(\gr, |\nabla G|^2\gr)-(G\gr, VG\gr).
\en
Let $\epsilon$ satisfy
$$\eb-V_\infty-\epsilon>0.$$
There exists $R$ such that
$$|V_-(x)-V_\infty|\leq \epsilon,\ \ \ |x|>R.$$
Let
$$a:=\sup_{|x|<R}G^2(x) V_-(x),\ \ \
b:=\frac{1}{2m} \sup_{x\in\BR} |\nabla G(x)|^2 .$$
Then
$$
\eb
\|G\gr\|^2
\leq b \|\gr\|^2
+
a\|\gr\|^2+
(V_\infty+\epsilon)\|G\gr\|^2,
$$
and
\eq{443}
\|G\gr\|^2\leq \frac{a+b}{\eb-V_\infty-\epsilon}\|\gr\|^2.
\en

(2) Suppose that  $G\in C^\infty(\BR)$.
Let $\chi\in C^\infty(\BR)$ be
$$\chi(x):=\lkk\begin{array}{ll}
1,&|x|\leq 1,\\
\theta(x), &1<|x|<2,\\
0,&|x|\geq 2,
\end{array}
\right.$$
such that
$\partial\theta(x)/\partial x_\mu \leq 0$, $\mu=1,2,3$,
$\sup_{x\in\BR}|\theta(x)|\leq 1$ and
$\sup_{x\in\BR} |\nabla \chi(x)|\leq c$ with some constant $c$.
Define
$\chi_n(x):=\chi(x/n^l)$ and set
$$G_n:=\chi_nG.$$
Note that
\eq{445}
|\nabla \chi_n(x)|=  \lkk\begin{array}{ll}0,& |x|< n^l,\\
\leq c n^{-l},& n^l\leq |x|\leq 2n^l,\\
0,& |x|> 2n^l.
\end{array}\right.
\en
It follows that by \kak{441},
$$\eb\|G_n\gr\|^2 \leq \frac{1}{2m}(\gr, |\nabla G_n|^2 \gr)+(G_n\gr,
V_-G_n\gr).$$
Since
$$
|\nabla G_n |^2 =
|\nabla \chi_n\cdot G_n|^2+|\chi_n\cdot \nabla G_n|^2+
2 (\nabla \chi_n\cdot G_n) \cdot (\chi_n\cdot \nabla G_n),$$
we have by \kak{445} and assumption (iv) of (2), for sufficiently large $n$,
\begin{eqnarray}
&&\frac{1}{2m}
(\gr, |\nabla G_n|^2\gr)\non \\
&&\leq
\frac{1}{2m}
\frac{c^2}{n^{2l}}\int_{n^l\leq |x|\leq 2n^l}|G(x)|^2\|\gr(x)\|_\fff^2 dx+
\frac{1}{2m}  (\gr, |\nabla G|^2\gr)\non \\
&&+
\frac{1}{m}  \lk
\frac{c^2}{n^{2l}}\int_{n^l\leq |x|\leq 2n^l}|G(x)|^2\|\gr(x)\|_\fff^2
dx\rk^\han
(\gr, |\nabla G|^2\gr)^\han \non \\
&&\label{lpo}
\leq \lk (2c^2d/m) +b  +\sqrt {8c^2db/m}  \rk \|\gr\|^2.
\end{eqnarray}
Moreover
$$\sup_{|x|<R}G_n^2(x)|V(x)|\leq \sup_{|x|<R}G^2(x)|V(x)|=a.$$
Thus we obtain that
\eq{442}
(\eb-V_\infty-\epsilon) \|G_n\gr\|^2
\leq
\lk b+a+(2c^2d/m) + \sqrt{8c^2db/m}\rk  \|\gr\|^2.
\en
Since $G_n$ is monotonously increasing in $n$ and $\lim_{n\rightarrow
\infty }G_n(x)=G(x)$,
by \kak{442} and the Lebesgue monotone convergence theorem, we see that
$ G \|\gr(\cdot)\|_\fff\in \LR$ and
$$\lim_{n\rightarrow \infty}\|G_n\gr\|^2=
\lim_{n\rightarrow \infty}\int _\BR G_n(x)^2 \|\gr(x)\|_\fff^2dx =
\|G\gr\|^2.$$
In particular by \kak{lpo},
$$\limn \frac{1}{2m} (\gr, |\nabla G_n|^2\gr)\leq b \|\gr\|^2. $$
Thus \kak{443} follows for $G\in C^\infty$.

(3) Suppose that $G\in C^1(\BR)$.
Let $\rho\in C_0^\infty(\BR)$ and $\rho>0$ such that
$\int_\BR\rho(x)dx=1$. Set $\re=\rho(\cdot/\epsilon)\epsilon^{-3}$.
Define
$$G_\epp:=\re\ast G.$$
Since $G_\epsilon\in C^\infty(\BR)$,
we have
$$
\|G_\epsilon \gr\|^2\leq
\frac{a_\epsilon+b_\epsilon}{\eb-V_\infty-\epsilon}  \|\gr\|^2,$$
where
$a_\epp$ and $b_\epp$ are $a$ and $b$ with $G$ replaced by $G_\epp$,
respectively.
Note that
$$\sup_{|x|<R}G_\epp(x)=
\sup_{|x|<R}\int_\BR \rho(z) G(x-\epp z) dz\leq
\sup_{|x|<R'}G(x),$$
with some $R'$,
and
$$
\sup_{x\in\BR}
|\nabla G_\epp(x)|
=
\sup_{x\in\BR}
\left | \int_\BR \rho(z) \nabla G(x-\epp z) dz\right|\leq
\sup_{x\in\BR} | \nabla G(x)|,
$$
which yields that
$$a_\epsilon\leq
\sup_{|x|<R'}G^2(x)|V(x)|:=a',\ \ \
b_\epp\leq b.$$
Hence
we obtain
that
$$
\|G_\epp \gr\|^2
\leq
\lk \frac{a'+b}{\eb -V_\infty-\epsilon}\rk \|\gr\|^2.
$$
From this,
$$\liminf_{\epp\rightarrow 0}
\|G_\epp \gr\|^2<\infty$$
follows.
Since
$$\liminf_{\epp\rightarrow 0}
G_\epp ^2(x) \|\gr(x)\|_\fff^2= G^2(x) \|\gr(x)\|_\fff^2,$$
Fatou's lemma yields that
$G\|\gr(\cdot)\|_\fff\in \LR$ and
$$
\|G \gr\|^2\leq
\liminf_{\epp\rightarrow 0} \|G_\epp \gr\|^2
\leq
\lk \frac{a'+b}{\eb -V_\infty-\epsilon}\rk \|\gr\|^2.
$$
Then
$$
\frac{\|G\gr\|^2}{\|\gr\|^2}
\leq \frac{a'+b}{\eb -V_\infty-\epsilon}
$$
follows. The proof is complete.
\qed
\begin{remark}
It can be  also proven that
$\gr\in D(e^{\beta |x|}\otimes 1)$ for some $\beta$
under some conditions on $V$, e.g., $V(x)=-e^2/|x|$. See \cite{gll,hhs}.
\end{remark}

\subsection{Proof of Lemma \ref{hi0}}
\label{gu}
{\it Proof of Lemma \ref{hi0}}\\
We prove  the lemma for $\mu=1$.
Proofs for $\mu=2,3$ are the same  as that of $\mu=1$.
Let $\epsilon=(a,0,0)\in\BD$.
We see that
$e^{i\epsilon x}$ maps $D(\pff)$ onto itself.
Thus
$$e^{i\epsilon x}\pf \Psi=\pf e^{i\epsilon x}\Psi+(\pf(\epsilon)-\pf)
e^{i\epsilon x}\Psi,\ \ \ \Psi\in D(\pff),$$
where
\begin{eqnarray*}
&&\pf(\epsilon):=\frac{1}{2m}(p\otimes 1+\epsilon-e\av)+V\otimes 1+1\otimes
\hf
-\frac{1}{2m}(\sigma\otimes 1)\cdot  \bv\\
&&=\pf+\frac{1}{m} (p\otimes 1-e\av)\cdot \epsilon
+\frac{1}{2m}|\epsilon|^2.
\end{eqnarray*}
We have
$$\lk\frac{e^{i\epsilon x}-1}{a}\rk
\pf \Psi= \pf\lk\frac{e^{i\epsilon x}-1}{a}\rk  \Psi
+\frac{1}{a}(\pf(\epsilon)-\pf) e^{i\epsilon x} \Psi.$$
By the assumption we see that
\eq{hi90}
s-\lim_{a\rightarrow 0} \lk\frac{e^{i\epsilon x}-1}{a}\rk
\pf\Psi=i(x_1\otimes 1) \pf\Psi.
\en
Directly we see that
\begin{eqnarray}
\frac{1}{a}(\pf(\epsilon)-\pf) e^{i\epsilon x} \Psi\non
&=&
\lk
\frac{1}{m} (p\otimes 1-e\av)_1+\frac{1}{2m}a\rk e^{i\epsilon x} \Psi\non \\
&=&
e^{i\epsilon x} \lk \frac{1}{m} (p\otimes
1-\epsilon -e\av)_1+\frac{1}{2m}a\rk \Psi\non \\
&&
\label{hi91}
\rightarrow \frac{1}{m}(p\otimes 1-e\av)_1\Psi
\end{eqnarray}
strongly as $a\rightarrow 0$.
From \kak{hi90} and \kak{hi91} it follows that
$$
s-\lim_{\epsilon\rightarrow 0}\pf\lk\frac{e^{i\epsilon x}-1}{a}\rk\Psi=i
(x_1\otimes 1)
\Psi-\frac{1}{m}(p\otimes 1-e\av)_1\Psi.$$
Since $\pf$ is closed and
$$s-\lim_{\epsilon\rightarrow 0}\lk\frac{e^{i\epsilon x}-1}{a}\rk\Psi
=i(x_1\otimes 1) \Psi,$$
it follows that $(x_1\otimes 1)\Psi\in D(\pf)$.  Thus (1) of Lemma \ref{hi0}
is proven.
Let
$$\Psi\in {\cal S}:=
C_0^\infty(\BR;\CC^2)\otimes [\bigoplus_{n=0}^\infty \otimes_s^n
C_0^\infty(\BR)].$$
Then we have
\eq{hi92}
[(x_1\otimes 1),   \pf]\Psi=\frac{i}{m}(p\otimes 1-e\av)_1\Psi.
\en
Hence
$$\|[(x_1\otimes1), \pf]\Psi\|\leq c \|(H_0+1)^\han\Psi\|$$
follows with some constant $c$.
Then
the closure $\ov{[(x_1\otimes 1), \pf]\lceil_{{\cal S}}}$ is well defined on

$D(H_0^\han)$ and satisfies
$$\ov{[(x_1\otimes 1), \pf]}\Psi=\frac{i}{m}(p\otimes 1-e\av)_1\Psi$$
for $\Psi\in D(H_0^\han)$.
In particular,
$$[(x_1\otimes1) ,\pf]\gr=
\ov{[(x_1\otimes1) ,\pf]\lceil_{{\cal S}}}\gr=\frac{i}{m}(p\otimes
1-e\av)_1\gr$$
follows, since $\gr\in D((x_1\otimes1) \pf)\cap D(\pf (x_1\otimes1) )$.
\qed

\footnotesize
{\bf Acknowledgements}
I am very grateful to A. Arai and M. Hirokawa for  useful comments and
discussions.


{\footnotesize
\begin{thebibliography}{99}

\bibitem{al3} S.  Albeverio,
An introduction to some mathematical aspects of
scattering theory in models of quantum fields,
{\it Scattering theory in mathematical physics}
 ed.  by    J.   A   LaVita  and  J.   P.   Marchand  (1974),
299--381.

\bibitem{ar14}
A.  Arai,   Essential spectrum of a self-adjoint operator on an
abstract Hilbert space of Fock type and applications to quantum field
Hamiltonians,   {\it J.   Math.   Anal.   Appl.  } {\bf 246}  (2000),
189--216.

\bibitem{ah1}
A.  Arai and M.  Hirokawa,
On the existence and uniqueness of  ground states of
a generalized spin-boson model,
{\it J.  ~Funct.  ~Anal.  } {\bf 151}  (1997),   455--503.


\bibitem{ah2}
A.  Arai and M.  Hirokawa,
Ground states of a general class of quantum field Hamiltonians,
{\it Rev. Math. Phys.} {\bf 12} (2000), 1085--1135.






\bibitem{ahh}
A.  Arai,   M.  Hirokawa,   and  F.  Hiroshima,
On the absence of eigenvectors of Hamiltonians in a class of massless
quantum field models without infrared cutoff,
{\it J.   Funct.   Anal.  } {\bf 168}  (1999),  470--497.
\bibitem{ahh2}
A.  Arai,   M.  Hirokawa,   and  F.  Hiroshima,
Regularity of ground states in quantum field models, in preparation.

\bibitem{ak}  A. Arai and H.  Kawano,
Enhanced binding in a general class of quantum field models,
{\it Rev. Math. Phys.} {\bf 15} (2003), 387--423.


\bibitem{bfs1}
V.  Bach,   J.  Fr\"ohlich,   I.  M.  Sigal,
Quantum electrodynamics of confined nonrelativistic particles,
{\it Adv.   Math.  } {\bf 137}  (1998),   299--395.
\bibitem{bfs3}
V.  Bach,   J.  Fr\"ohlich,   I.  M.  Sigal,
Spectral analysis for systems of atoms and molecules coupled to the
quantized radiation field,   {\it Commun.   Math.   Phys.  } {\bf 207}
(1999),   249--290.


\bibitem{bcv}
J. M.  Barbaroux, T.  Chen and S. Vugalter,
Binding conditions for atomic N-electron systems in non-relativistic QED
arXiv math-ph/0304019, preprint, 2003.


\bibitem{bdg3}
J.  M.  Barbaroux, M. Dimassi and J. C. Guillot, Quantum electrodynamics of
relativistic bound states with cutoffs, arXiv:math-ph/0312011, preprint
2003.




\bibitem{fr2}
J.  Fr\"ohlich,   Existence of dressed one electron states in a class of persistent models,   {\it Fortschritte der Physik} {\bf 22}  (1974),   159--198.   



\bibitem{fgs}
J.  Fr\"ohlich,  M.  Griesemer and B.  Schlein,
Asymptotic electromagnetic fields in a mode of quantum-mechanical
matter interacting with the quantum radiation field,
{\it Adv. in Math.} {\bf 164} (2001), 349--398.

\bibitem{ge}
C.  G\'erard,
On the existence of ground states for massless Pauli-Fierz Hamiltonians,
{\it Ann.  Henri  Poincar\'e} {\bf 1}  (2000),  443--459.


\bibitem{gll}
M.  Griesemer,   E.  Lieb and M.  Loss,
Ground states in non-relativistic quantum electrodynamics,
{\it Invent. Math.} {\bf 145} (2001), 557--595.


\bibitem{gr0}
L.  Gross,
A noncommutative extension of the Perron-Frobenius theorem,  {\it Bull.
Amer.  Math.  Soc. } {\bf 77}  (1971),  343--347.

\bibitem{gr1}
L.  Gross,   Existence and uniqueness of physical ground states,
{\it J.   Funct.   Anal.  } {\bf 10}   (1972),   52--109.

\bibitem{ho1}
R.  H\o egh-Krohn,
Asymptotic fields in some models of quantum field theory I,
{\it J.   Math.   Phys.  } { \bf  9}  (1968),   2075--2080.




\bibitem{hiro0}
M.  Hirokawa
Infrared Catastrophe for Nelson's Model, mp-arc 03-512,  preprint, 2003.

\bibitem{hiro}
M.  Hirokawa,
Mathematical Addendum for
``Infrared Catastrophe for Nelson's Model" (mp-arc 03-512),
mp-arc,03-551, preprint, 2003.

\bibitem{hhs}M. Hirokawa, F. Hiroshima, and H. Spohn,
 Ground state for point particles interacting through a massless scalar Bose
 field,
arXiv:math-ph/0211050, preprint, 2002, to be published in {\it Adv. Math.}





\bibitem{h12}
F.  Hiroshima,
Ground states and spectrum of quantum electrodynamics of
non-relativistic particles,
{\it Trans.   Amer.   Math.   Soc.  } {\bf 353} (2001), 4497--4528.
\bibitem{h8}
F.  Hiroshima,
Ground states of a model in nonrelativistic quantum electrodynamics I,
{\it J.   Math.   Phys.  } {\bf 40}  (1999),   6209--6222.
\bibitem{h9}
F.  Hiroshima,
Ground states of a model in nonrelativistic quantum electrodynamics II,
{\it J.    Math.   Phys.  }  {\bf 41}   (2000),   661--674.


\bibitem{h11}
F.  Hiroshima,
Essential self-adjointness of translation-invariant quantum field models for
arbitrary coupling constants,
{\it Commun.   Math.   Phys.  } {\bf 211}   (2000),   585--613.


\bibitem{h16}
F.  Hiroshima,
Self-adjointness
of the Pauli-Fierz Hamiltonian for arbitrary values of coupling constants,
{\it Ann. Henri  Poincar\'e}, {\bf 3} (2002), 171--201.


\bibitem{hisp2}
F.  Hiroshima and H.  Spohn,
Ground state degeneracy of the Pauli-Fierz model with   spin,
{\it Adv. Theor. Math. Phys.} {\bf 5} (2001), 1091--1104.


\bibitem{kato}
T.  Kato,   {\it Perturbation  Theory  for  Linear  Operators},    Springer-Verlag,   
1966.   

\bibitem{gl1}
E.  Lieb and M.  Loss,   Existence of atoms and molecules in
non-relativistic quantum electrodynamics, {\it Adv. Theor. Math. Phys.} {\bf
7} (2003),  667-710.



\bibitem{gl2}
E.  Lieb and M.  Loss,   A note on polarization vectors in quantum
electrodynamics,
arXiv:math-ph/0401016, preprint, 2004.



\bibitem{ne3}
E.  Nelson,   Interaction of nonrelativistic particles with a quantized
scalar field,   {\it J.   Math.   Phys.  }{\bf  5}  (1964),   1190--1197.


\bibitem{rs1}
M.  Reed and B.  Simon,   {\it Methods of Modern Mathematical Physics I},
Academic Press,  1980.
\bibitem{rs2}
M.  Reed and B.  Simon,   {\it Methods of Modern Mathematical Physics II},
Academic Press,  1975.
\bibitem{rs3}
M.  Reed and B.  Simon,   {\it Methods of Modern Mathematical Physics III},
Academic Press,  1979.

\bibitem{rs4}
M.  Reed and B.  Simon,   {\it Methods of Modern Mathematical Physics IV},
Academic Press,   1978.

\bibitem{sl}
A. D. Sloan, The polaron without cutoffs in two space dimensions, 
{\it J. Math. Phys.} {\bf 15} (1974), 190--201.

\bibitem{sp}
H.  Spohn,
Ground state of quantum particle coupled to a scalar boson field,
{\it  Lett.   Math.   Phys.  } {\bf 44}   (1998),   9--16.

\bibitem{wei}
J. Weidmann, {\it Linear Operators in Hilbert Spaces}, GTM 68,
Springer-Verlag, 1980.
\end{thebibliography}
}
\end{document}